\documentclass[fleqn,usenatbib]{mnras}

\usepackage{newtxtext}%,newtxmath}

\usepackage[T1]{fontenc}

\DeclareRobustCommand{\VAN}[3]{#2}
\let\VANthebibliography\thebibliography
\def\thebibliography{\DeclareRobustCommand{\VAN}[3]{##3}\VANthebibliography}

\usepackage{graphicx}	% Including figure files
\usepackage{amsmath}	% Advanced maths commands
\usepackage{amssymb}	% Extra maths symbols
\usepackage{subfigure}
\usepackage[usenames,dvipsnames]{xcolor}
\usepackage{ulem}
\newcommand{\msun}{{\rm M_{\sun }}}
\newcommand{\Mo}{\msun}
\usepackage{epstopdf}

\title[Nuclear star cluster mergers]{Searching for clues of past binary supermassive black hole mergers in nuclear star clusters}

\author[A.~Mastrobuono-Battisti et al.]{Alessandra Mastrobuono-Battisti$^{1}$\thanks{E-mail:
alessandra.mastrobuono-battisti@obspm.fr},
Go Ogiya$^{2,3}$,
Oliver Hahn$^{4,5}$,
and Mathias Schultheis$^{6}$
\\
$^{1}$GEPI, Observatoire de Paris, PSL Research University, CNRS, Place Jules Janssen, 92190 Meudon, France\\
$^{2}$Institute for Astronomy, School of Physics, Zhejiang University, Hangzhou 310027, China\\
$^{3}$Waterloo Centre for Astrophysics, University of Waterloo, Waterloo, ON N2L 3G1, Canada\\
$^{4}$Department of Astrophysics, University of Vienna, Türkenschanzstraße 17, 1180, Vienna, Austria\\
$^{5}$Department of Mathematics, University of Vienna, Oskar-Morgenstern-Platz 1, 1090, Vienna, Austria\\
$^{6}$Université Côte d’Azur, Observatoire de la Côte d’Azur, CNRS, Laboratoire Lagrange, Nice, France \\
}

\date{Accepted XXX. Received YYY; in original form ZZZ}

\pubyear{2015}

\begin{document}
\label{firstpage}
\pagerange{\pageref{firstpage}--\pageref{lastpage}}
\maketitle

\begin{abstract}
Galaxy mergers are common processes in the Universe. As a large fraction of galaxies hosts at their centres a central supermassive black hole (SMBH), mergers can lead to the formation of a supermassive black hole binary (SMBHB). The formation of such a binary is more efficient when the SMBHs are embedded in a nuclear star cluster (NSC). NSCs are dense and massive stellar clusters present in the majority of the observed galaxies. Their central densities can reach up to $10^7\,\Mo/{\rm pc}^3$ and their masses can be as large as a few $10^7\,\Mo$. The direct detection of an SMBHB is observationally challenging. In this work, we illustrate how the large scale structural and dynamical properties of an NSC can help to identify nucleated galaxies that recently went through a merger that possibly led to the formation of a central SMBHB. Our models show that the merger can imprint signatures on the shape, density profile, rotation  and velocity structure of the NSC. The strength of the signatures depends on the mass ratio between the SMBHs and on the orbital initial conditions of the merger.  In addition, the number of hypervelocity stars produced in the mergers is linked to the SMBHB properties. The merger can also contribute to the formation of the nuclear stellar disc of the galaxy. 
\end{abstract}


%%%%%%%%%%%%%%%%%%%%%%%%%%%%%%%%%%%%%%%%%%%%%%%%%%

%%%%%%%%%%%%%%%%% BODY OF PAPER %%%%%%%%%%%%%%%%%%

\section{Introduction}
%%few lines on NSDs are still missing!!!
Nearly all galaxies with a stellar mass larger than $10^{10}\,\Mo$ host at their centres a supermassive black hole (SMBH) with a mass enclosed between $10^6\,\Mo$ and $10^9\,\Mo$ \citep[see e.g.][]{Ferrarese05, Kormendy13}. Galaxies with stellar masses between $10^8\,\Mo$ and $10^{10}\,\Mo$ have, instead, their centres dominated by dense and massive stellar systems, called nuclear star clusters \citep[NSCs,][]{Boeker04, Cote06, Boeker10, Neumayer11, Turner12, Georgiev14, denbrok14, SJ19, Neumayer20}. Typically, NSCs have masses of $10^6$-$10^7\,\Mo$ and are characterized by half-mass radii of a few parsecs.
Despite the links existing between their properties and those of their parent galaxy \citep{Rossa06, Ferrarese06, Wehner06, Neumayer20}, the NSC formation process is not yet clear; NSCs are thought to form through a mixture of in situ star formation \citep{Loose82, LE03, Milosavljevic04, Nayakshin05, Paumard06, Schinnerer06, Schinnerer08, Hobbs09, Mapelli12, Mastrobuono19} and dynamical friction-driven star cluster decay and mergers \citep{Tremaine75, CD93, Antonini12, Gnedin14, Mastrobuono14,PMB14, Arcasedda15, AN15, Tsatsi17, Abbate18}. 
In a fraction of observed galaxies, the NSC coexists with a central SMBH \citep{Neumayer12, Nguyen19}. One of these galaxies is the Milky Way, whose centre hosts both Sgr A*, our $4.3\times10^6\,\Mo$ SMBH, and a surrounding NSC of about $2.5\times10^{7}\,\Mo$ \citep{GH98, EI05, GI09, BGS16, GP17, Schoedel14b}.\\
Mergers between galaxies are common events in the Universe. In the $\Lambda$-cold dark matter ($\Lambda$CDM) scenario, they are considered responsible for the mass growth of  galaxies \citep{deBlok10, Newman12, Hill17}. 
If the galaxies that merge have comparable masses (i.e., mass ratio larger than 0.1) and they both host a central SMBH, the merger will likely produce an initially gravitationally unbound SMBH pair, on the distance scale of $100$\,pc \citep{Volonteri03, Kazantzidis05, Callegari09}. 
When the separation between the components of the pair decreases down to a radius that encloses the total mass of the two black holes, the system becomes gravitationally bound, leading to the formation of an SMBH binary (SMBHB).
If the  two merging galaxies are nucleated, i.e., if they host a central NSC that embeds the SMBH, the separation between the SMBHs decreases more efficiently, quickly forming an SMBHB \citep{VanW2014, Biava19, Ogiya20}. This process has strong implications in the generation of gravitational waves (GWs) detectable by the future Laser Interferometer Space Antenna, LISA \citep{AS17}, and by the ongoing International Pulsar Timing Array experiment \citep{Hobbs10}. We note that the coalescence of SMBHs more massive than $10^8\,\Mo$ and the following transfer of energy to the surrounding stars can result in the effective disruption of the NSC. This mechanism is considered to be the cause of the absence of NSCs in  galaxies that host massive SMBHs \citep{Quinlan97, Cote06, Milosavljevic01, Neumayer20}. \\
\cite{VanW2014} studied the orbital evolution of SMBHs in merging galaxies using hydrodynamical simulations with a resolution of the order of $10\,$pc. The results of these simulations show that the merger between star bursting nuclei can lead to the disruption of one of the two nuclei and to the formation of a central cusp in the other nucleus. These are processes that significantly shorten the timescale for the formation of the SMBHB.\\
\cite{Ogiya20} recently performed high-resolution direct $N$-body simulations of the merger between two NSCs, each containing a central SMBH. The mergers considered by the authors are expected to happen during galaxy major mergers, i.e. mergers between galaxies of comparable masses. During the merger, dynamical effects such as dynamical friction, stellar hardening, and the extra deceleration force provided by the so-called `ouroboros effect' cooperate to efficiently reduce the separation between the SMBHs. In all the explored merger cases, the binary becomes hard and the two SMBHs coalesce in less than a Hubble time, leading to the emission of GWs.\\
The results summarized above imply that SMBHBs should be commonly present at the centre of galaxies. However, despite  extensive surveys, only a few  such candidates have been identified  \citep{Rodriguez06, Burke11, Tremblay16,Millon22}, suggesting that SMBHBs quickly merge or escape their galactic nucleus \citep{Merritt05}.\\
Nonetheless, the fact that some candidates have been actually observed indicates that several others might exist. The small number of current detections might be, therefore, due to observational biases. Although difficult to  detect on small spatial scales, SMBHBs could be indirectly detected, looking for the dynamical and structural signatures left on the surrounding NSC by the merger event that led to their formation. These properties  might be observable and can help to direct SMBHB searches, as we detail in this work. \\
Another helpful tool to identify SMBHBs is through hypervelocity stars (HVSs). HVSs are stars ejected from galactic nuclei with velocities equal to or larger than 1000\,km/s \citep{Hills88, Yu03, Brown15}. These stars can be produced both through interactions with a single or a binary SMBH\footnote{Mechanisms that can produce HVSs include the ejection of one of the stars forming a stellar binary during a close encounter with a central SMBH \citep{Hills88}, the ejection of a single star by a hard SMBHB that, following this interaction, becomes harder \citep{Yu03} and the ejection of a star bound to an SMBH due to the interaction with a second SMBH \citep{Gualandris05,Guillard16}.} and, consequently, their properties have been used to investigate the presence and characteristics of SMBHs in galactic nuclei \citep{Yu03, Darbha19}. Noticeably, they have been also used to explore the star formation history \citep{Kollmeier07} and to constrain the dark matter and baryonic gravitational potential \citep{Gnedin05, Kenyon14, Rossi17} of their parent galaxy. \\
In this paper, we analyse the NSC merger simulations run by \cite{Ogiya20} to study the signatures left by this process on the structure of the final central cluster, depending on the orbital initial conditions of the progenitors and on the mass ratio between the central SMBHs. The age of the analysed systems is in all cases 20\,Myr, a time at which the binary has hardened significantly, slowing down the simulation. \\ 
Galaxies that host NSCs of mass similar to what is considered in this work ($10^7\,\Mo$) span a large range of stellar masses and dynamical properties. The large spread existing in the relationships found between the galaxy stellar mass, the NSC and SMBH mass imply that part of the nucleated galaxies with stellar masses between $10^8\,\Mo$ and $10^{10}\,\Mo$ might not contain a central SMBH \citep{Neumayer20}. Therefore, we compare our results with two new merger simulations run either with two SMBH-less NSCs or with only one of the two merging NSCs hosting a central SMBH, so to clearly identify the dynamical effects of the SMBHB.\\
The paper is structured as follows. In Section \ref{sec:sims} we summarize the models and describe the simulations. In Section \ref{sec:res} we present the results of our analysis. In Section \ref{sec:disc} we discuss our results and draw our conclusions.
%%%mention NSD results!!!

%%%%%%%%%%%%%%%%%%%%%%%%%%%%%%%%%%%%%%%%%%%%%
\begin{table*}
    \centering
        \begin{tabular}{c c c c c c c c}        % centered columns (4 columns)
        \hline\hline 
        ID: property & $q$ & $d_i$ (pc) & $\eta$ & $r_h$ (pc) & $M_{tot}\,(\Mo)$ & $M_{d_i}\,(\Mo)$ & $M_1/{M_{d_i}}$\\
        \hline
        M1: small-$q$ & 0.01 & 20 & 1.0 & 9.65 & $1.9\times10^7$ & $1.4\times10^7$ & 0.5\\
        M2: small-$\eta$ & 0.1 & 20 & 0.5  & 10.0 & $1.9\times10^7$ & $1.4\times10^7$ & 0.5\\
        M3: fiducial & 0.1 & 20 & 1.0 &  11.1 & $1.9\times10^7$ & $1.3\times10^7$ & 0.5\\
        M4: large-$d_i$ & 0.1 & 50 & 1.0 &  12.2 & $1.8\times10^7$ & $1.5\times10^7$ & 0.5\\
        M5: large-$q$ & 1.0 & 20 & 1.0 &  13.5 & $1.8\times10^7$ & $1.2\times10^7$ & 0.5\\
         \hline
    \end{tabular}
    \caption{Summary of the properties of the simulations with two NSCs, each hosting a central SMBH. The table lists the name and main defining property of each model (ID: property), the SMBH mass ratio ($q$), the initial distance $d_i$ between the SMBHs, the parameter $\eta$ which quantifies the initial relative velocity between the NSCs, the half-mass radius and the total mass of the final NSC ($r_h$ and $M_{tot}$), the mass of the final NSC within $d_i$ ($M_{d_i}$) and the ratio between the mass of the NSC hosting the more massive SMBH (or labelled as NSC1) and $M_{d_i}$ ($M_1/{M_{d_i}}$).}
    \label{tab:tab1}
\end{table*}

\begin{table*}
    \centering
        \begin{tabular}{c c c c c c c c c}        % centered columns (4 columns)
        \hline\hline 
        ID & $M_{\bullet,NSC1}\,(\Mo)$ & $d_i$ (pc) & $\eta$ & $r_h$ (pc) & $M_{tot}\,(\Mo)$ & $M_{d_i}\,(\Mo)$ & $M_1/{M_{d_i}}$\\
        \hline
        NO SMBH  & 0  & 20 & 1.0 &  8.0 & $2.0\times10^7$ & $1.5\times10^7$ & 0.5\\
        ONE SMBH & $10^6$ &  20 & 1.0 &  8.2 & $2.0\times10^7$ & $1.5\times10^7$ & 0.5\\

         \hline
    \end{tabular}
    \caption{Summary of the properties of the two additional models run without any SMBH or with only one SMBH. The table lists the name of the model (ID), the mass of the  SMBH ($M_{\bullet,NSC1}$), the initial distance $d_i$ between the centres of the two NSCs, the parameter $\eta$ which quantifies the initial relative velocity between the NSCs, the half-mass radius and the total mass of the final NSC ($r_h$ and $M_{tot}$), the mass of the final NSC within $d_i$ ($M_{d_i}$) and the ratio between the mass of the NSC hosting the more massive SMBH (or labelled as NSC1) and $M_{d_i}$ ($M_1/{M_{d_i}}$).}
    \label{tab:tab_app_comp}
\end{table*}
%%%%%%%%%%%%%%%%%%%%%%%%%%%%%%%%%%%%%%%%%%%%%

\section{Models and simulations}\label{sec:sims}
We analyse the five NSC merger simulations presented by \cite{Ogiya20} and two additional simulations, one run with only one NSC hosting a central SMBH and one with two SMBH-less NSCs. 
In these simulations, the merger between two NSCs leads to the formation of a new NSC. If both  NSCs contain a central SMBH the final NSC will then host an SMBHB. In the following paragraphs, we briefly summarize the adopted initial conditions and the characteristics of the code used to run the simulations \citep[more details can be found in][]{Ogiya20}. 

\subsection{Initial conditions and {\it N}-body code} \label{sec:ICs}
The main simulation set-up conditions and the properties of the final NSCs are summarized in Tables \ref{tab:tab1} and \ref{tab:tab_app_comp}.\\
The initial NSC spatial density distribution is modelled using a \cite{Dehnen93} profile
\begin{equation}
    \rho(r)=\frac{(3-\gamma)M_{NSC}}{4\pi}\frac{r_0}{r^\gamma(r+r_0)^{4-\gamma}}
\end{equation}
where $M_{NSC}$ is the total NSC mass, $r_0$ is its core radius and $\gamma$ is the slope of the profile in the inner regions of the NSC.
All the simulations assume $M_{NSC}=10^7\,\Mo$, a cored density profile with $\gamma=0$ and a core radius $r_0=1.4\,$pc. The value adopted for $r_0$ corresponds to a half-light radius of $4\,$pc which is a typical value for NSCs with masses of $10^7\,\Mo$ \citep{Neumayer20}. The effect of dynamical friction is less efficient in cored density distributions. Indeed, dynamical friction shrinks the SMBHB orbit more strongly in NSCs having $\gamma > 0$. As two progenitor NSCs are simulated, the total stellar mass in each run is $2 \times 10^7 \, \Mo$.
The mass of the most massive SMBH is always set to be equal to $10^6\,\Mo$ while the mass of the secondary SMBH is either $10^4\,\Mo$, $10^5\,\Mo$ or $10^6\,\Mo$, such as to represent the scatter observed among SMBH masses in NSCs \citep{Georgiev16}. Each NSC is modelled using 65\,536 $N$-body particles of the same mass; this choice corresponds to a mass resolution of $152.6\,\Mo$. 
The SMBH is introduced at the centre of the NSC with zero velocity and the velocities of the stellar particles are drawn using the \cite{Eddington16} formula, taking into account the presence of the SMBH in the calculation of the gravitational potential. For simplicity, we identify the NSC hosting the more massive (primary) SMBH as NSC1, and the NSC hosting the less massive (secondary) SMBH as NSC2. One of the additional comparison models that we have run has no central SMBH in either of the NSCs and the other one has only a central SMBH of $10^6\,\Mo$ SMBH, introduced inside NSC1 following the same procedure used for the two-SMBH models.\\
The initial separation between the two NSCs, $d_i$, is set to be either $20\,$pc or $50\,$pc, values that correspond to the initial distance between the central SMBHs. The initial distances are significantly larger than the NSCs' effective radii and ensure that the two SMBHs are not initially bound. The parameter  $\eta$, that is used to characterize their initial angular momentum, can be either 0.5 or 1.0 and is defined through the initial relative velocity between the NSCs 
\begin{equation}
    v_i = \eta\sqrt{\frac{GM_*(d_i)}{d_i}}
\end{equation}
where $M_*(d_i)$ is the mass of the merging systems calculated as the sum of the NSC masses enclosed within a distance $d_i/2$ from their centres. The value of the initial angular momentum also depends on $d_i$, with larger $d_i$ indicating a larger angular momentum.  The low eccentricity values chosen for the cluster relative orbits follow from the circularizing effect that the dynamical friction has on the decaying NSCs \citep{Penarubia2004}.  The primary SMBH is initially located at the origin of the reference frame with zero velocity, while the secondary is located on the $x$-axis with a total velocity $v_i$, oriented in the $y$ direction.  The centre of each component and that of the merged NSC is defined as its centre of density. Both the NO SMBH and the ONE SMBH models are run assuming $d_i=20$\,pc and $\eta=1.0$.

The simulations have been run with NBDODY6++GPU \citep{Wang15}, the GPU-parallelised version of NBODY6 \citep{Aarseth03} an effective and accurate direct $N$-body code for collisional dynamics.   All the input parameters necessary to reproduce the simulations with this code are described in \cite{Ogiya20}.

%%%
\begin{figure*}
    \raggedright
    \includegraphics[width=0.33\textwidth]{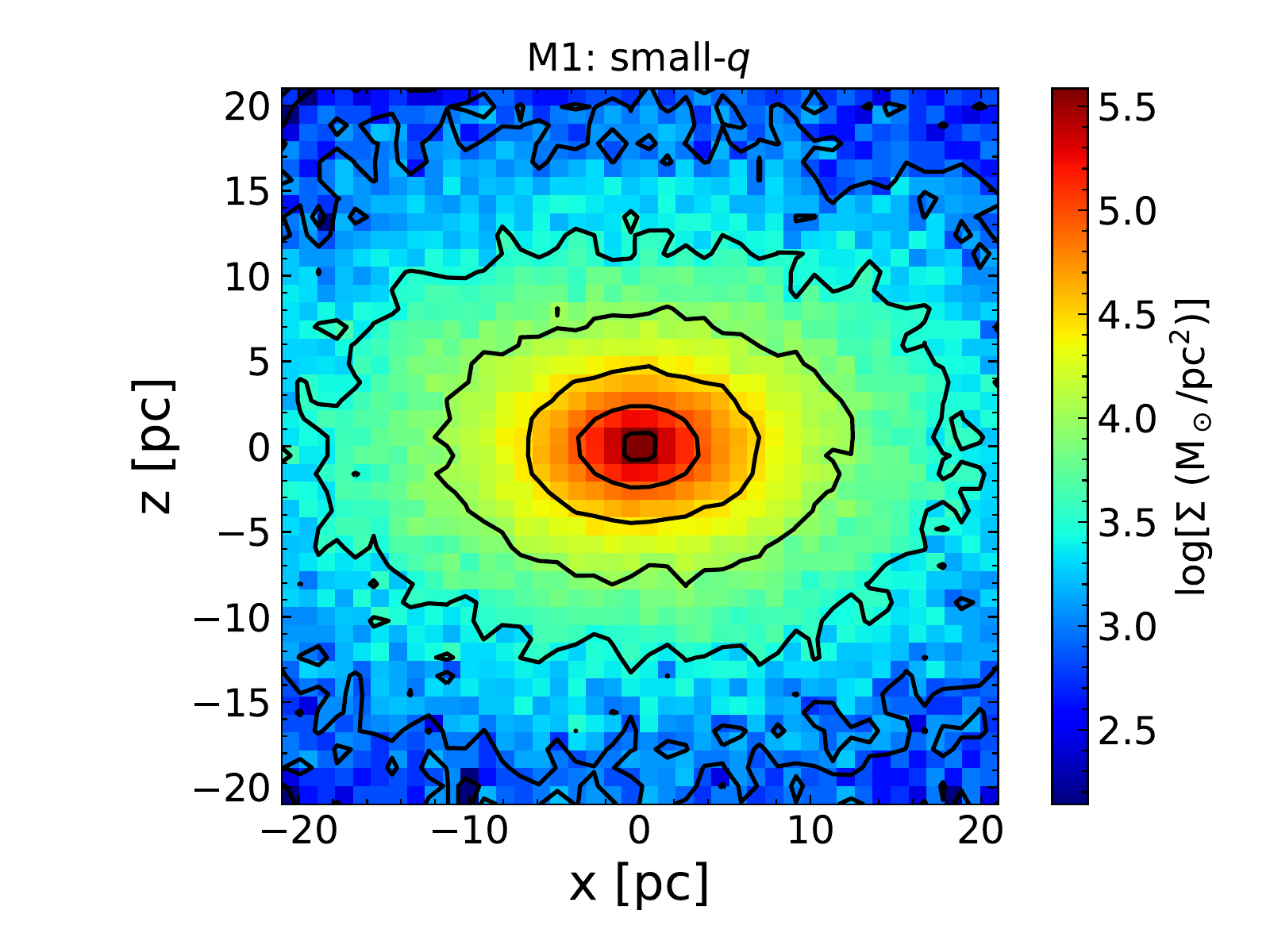}
    \includegraphics[width=0.33\textwidth]{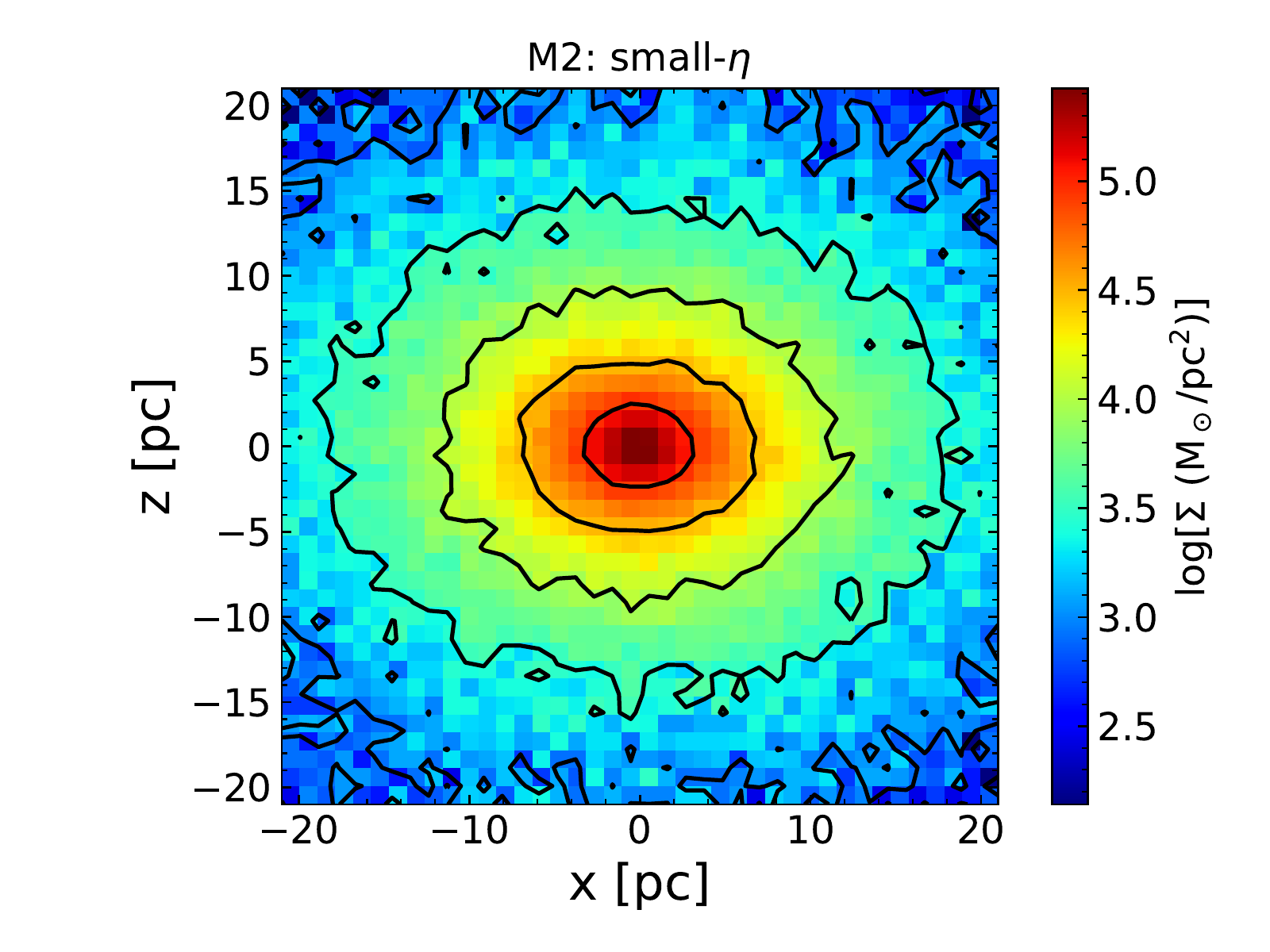}
    \includegraphics[width=0.33\textwidth]{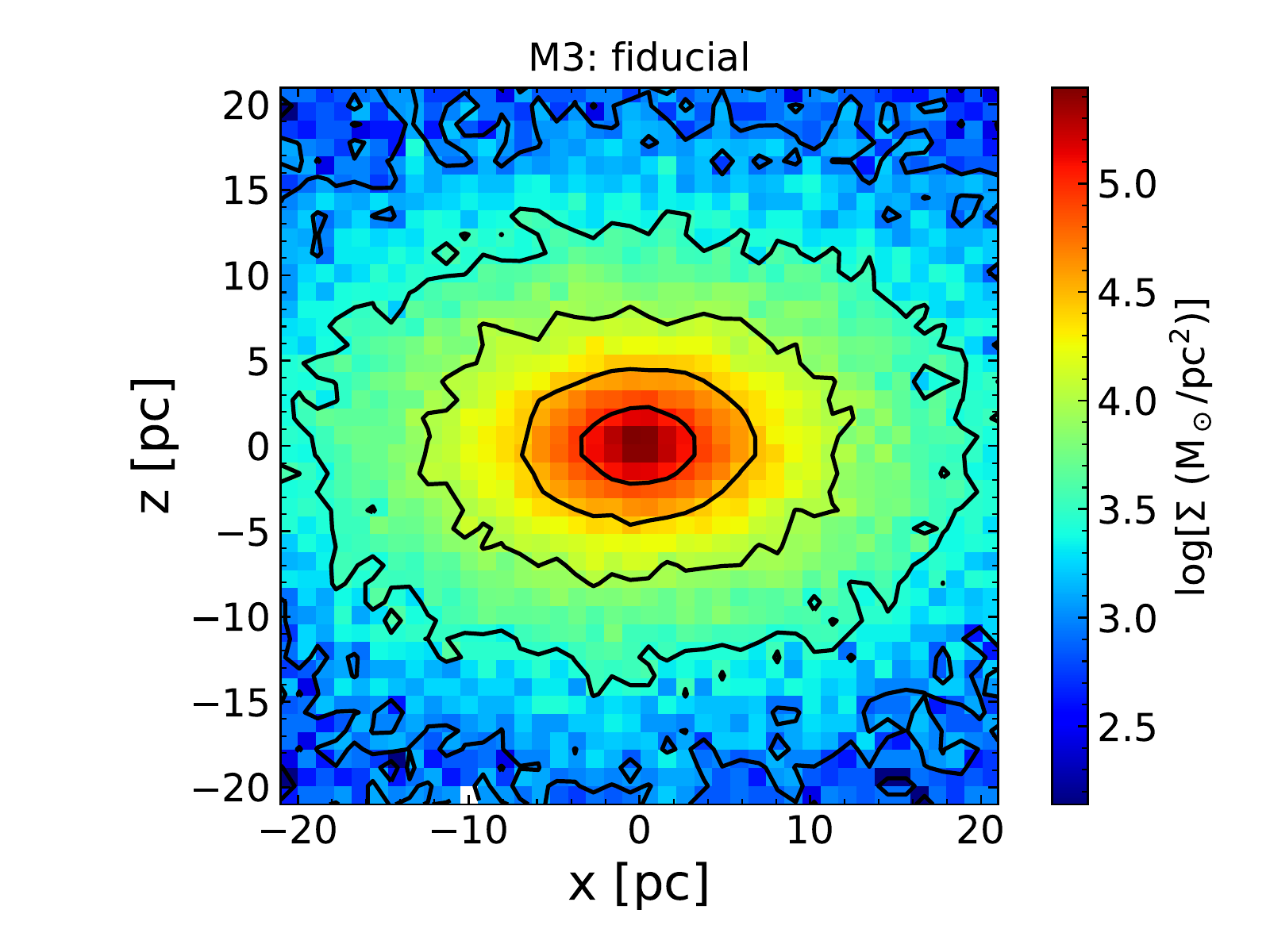}
    \includegraphics[width=0.33\textwidth]{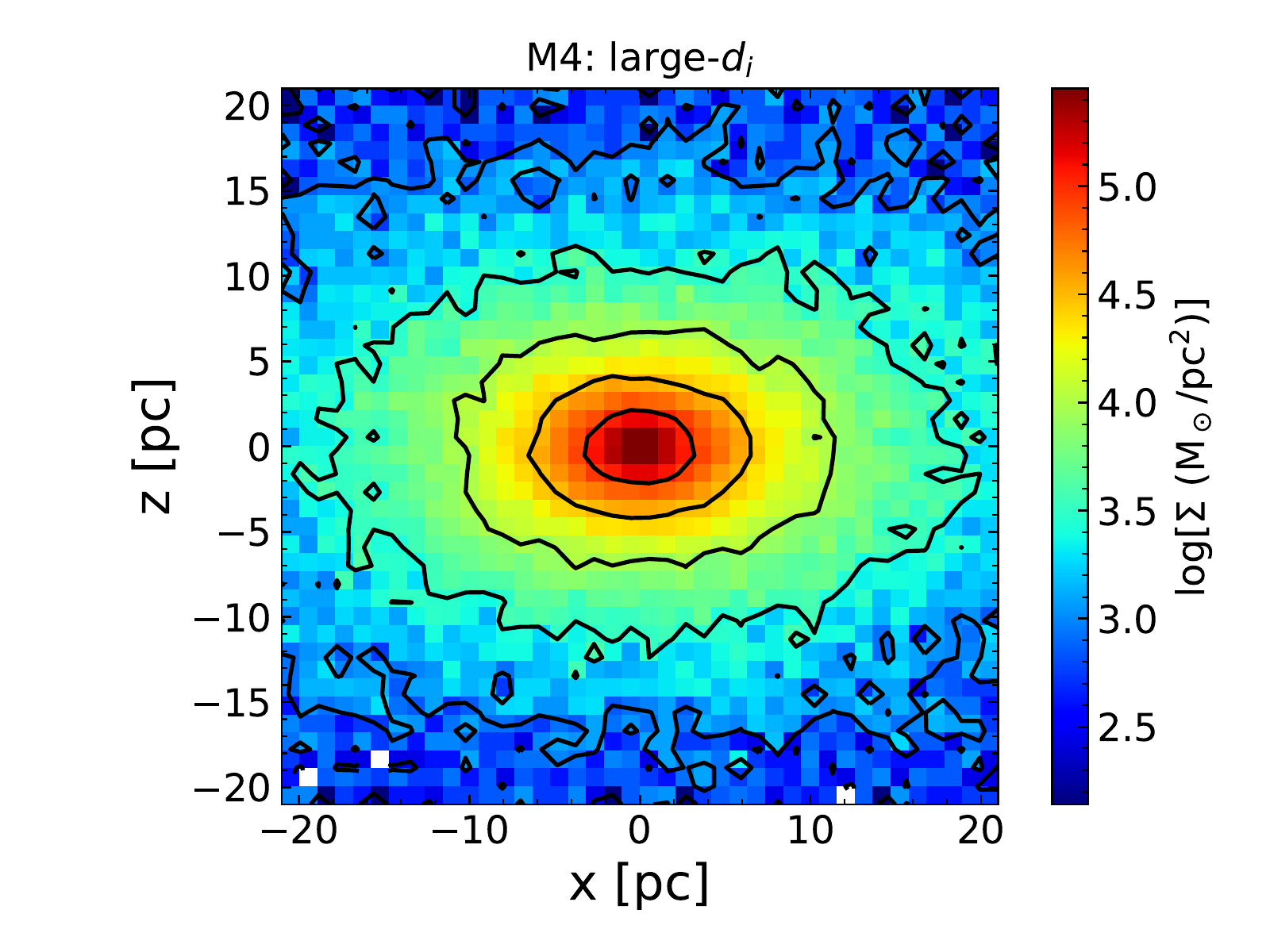}
    \includegraphics[width=0.33\textwidth]{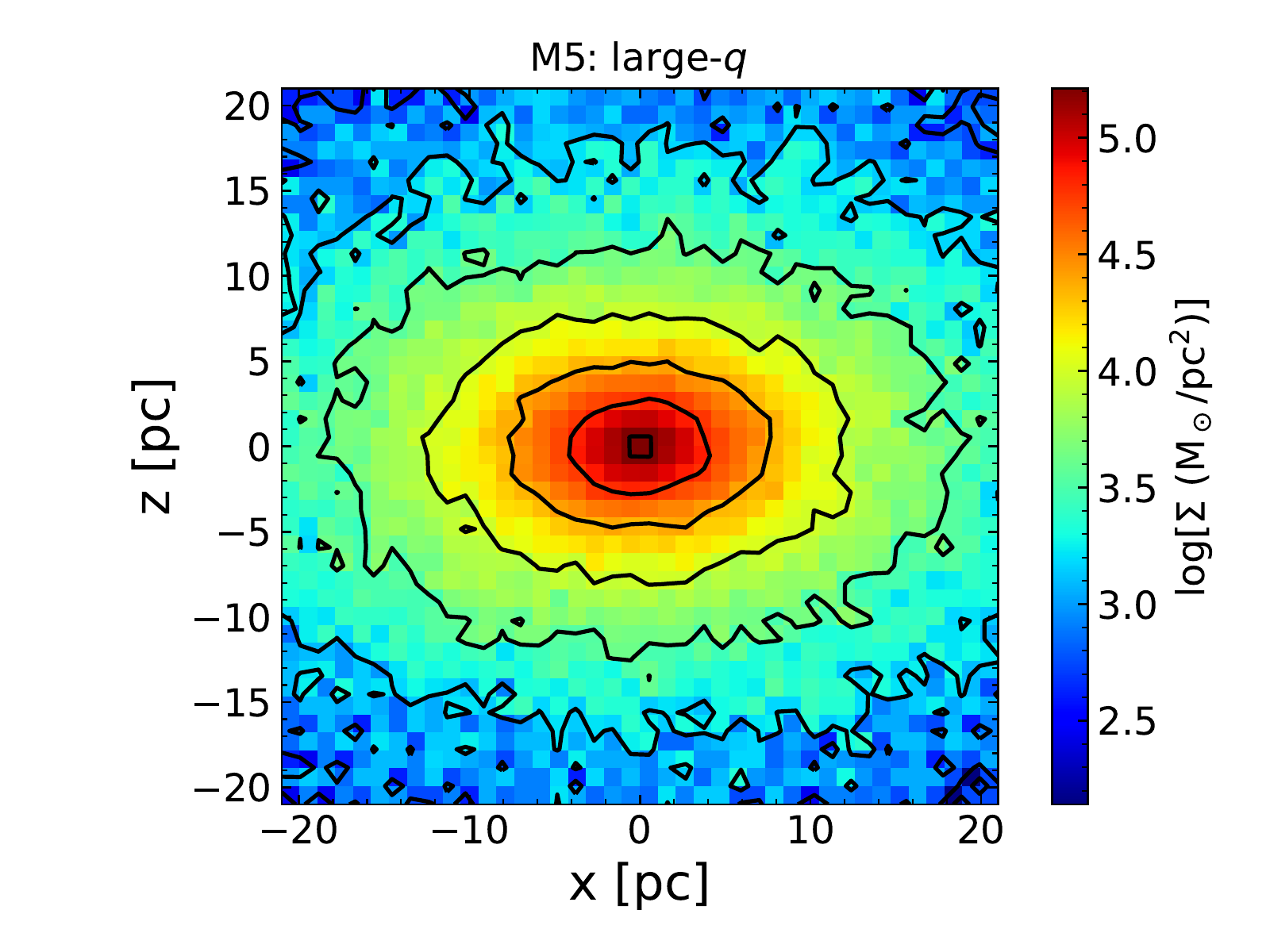}
    \includegraphics[width=0.33\textwidth]{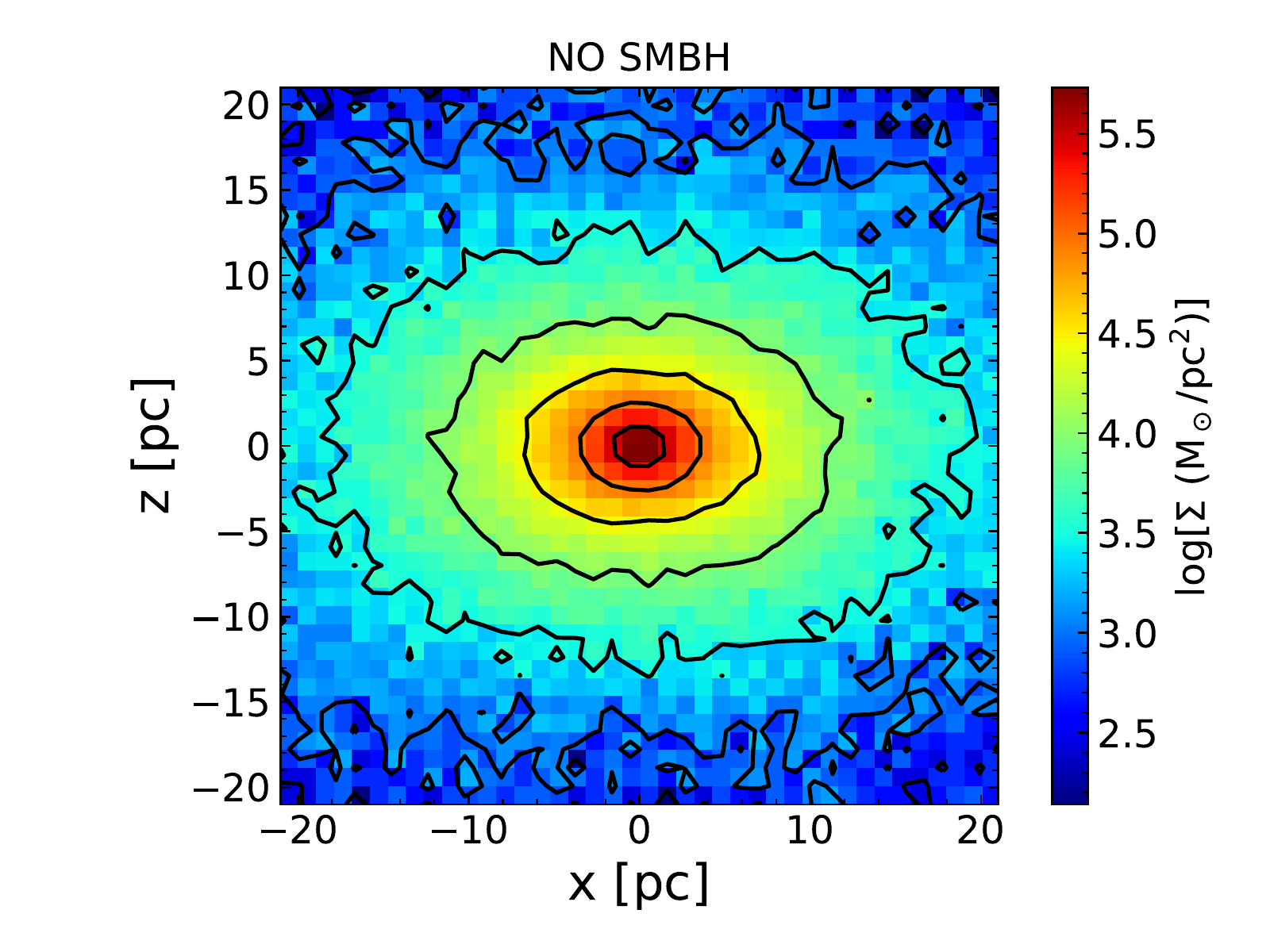}
    \includegraphics[width=0.33\textwidth]{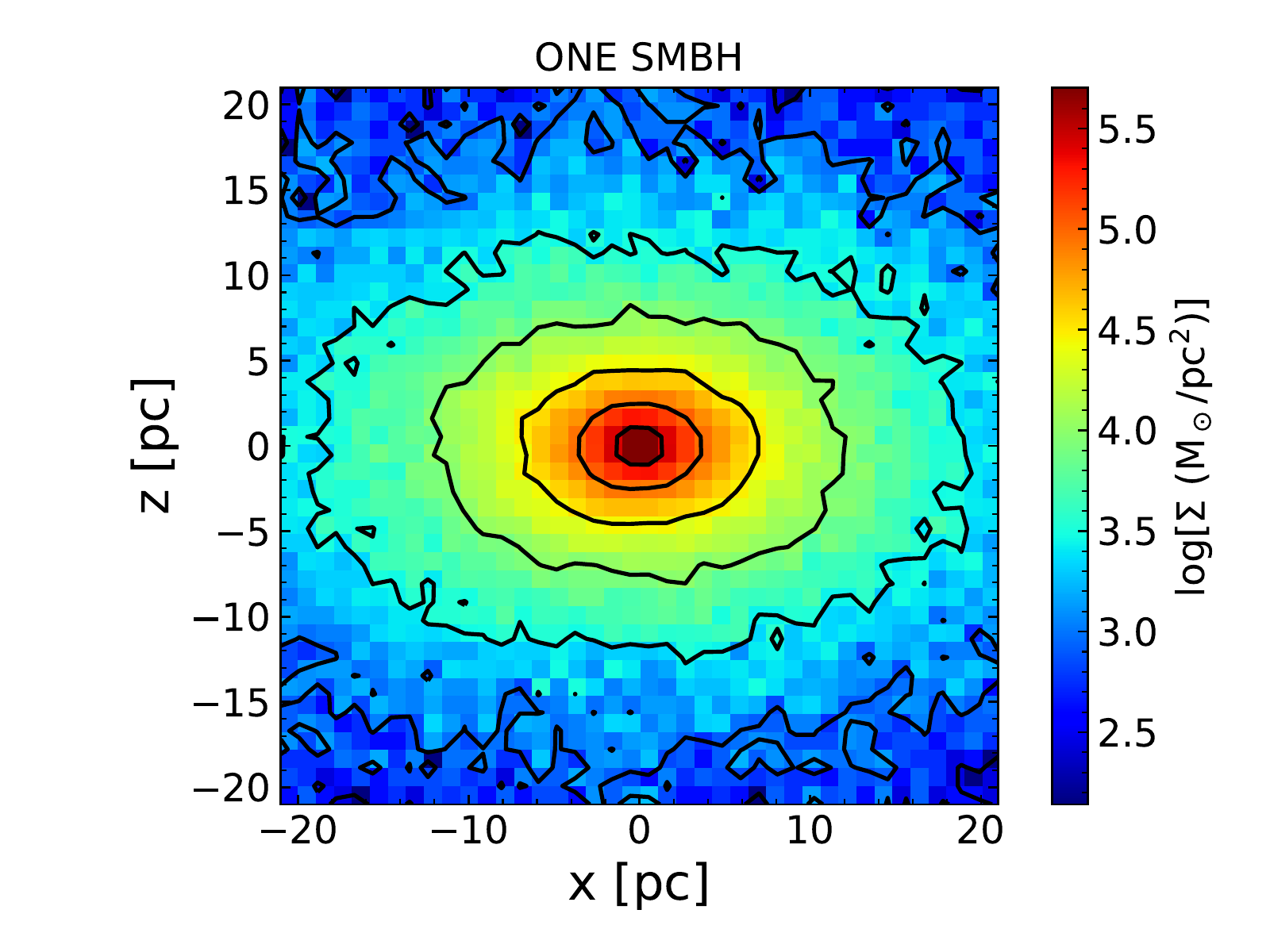}
    \caption{
    Density maps for the final NSCs obtained in the two-SMBH merger simulations and in the two additional models run in this work. From top to bottom and from left to right the maps are the model M1, M2, M3, M4, M5, NO SMBH and ONE SMBH. All the systems are seen edge-on and are clearly flattened and centrally concentrated. The structure of the NSCs primarily depends on the initial merger orbit of two NSCs.}
    \label{fig:map_xz}
\end{figure*}

\begin{figure*}
    \raggedright
    \includegraphics[width=0.33\textwidth]{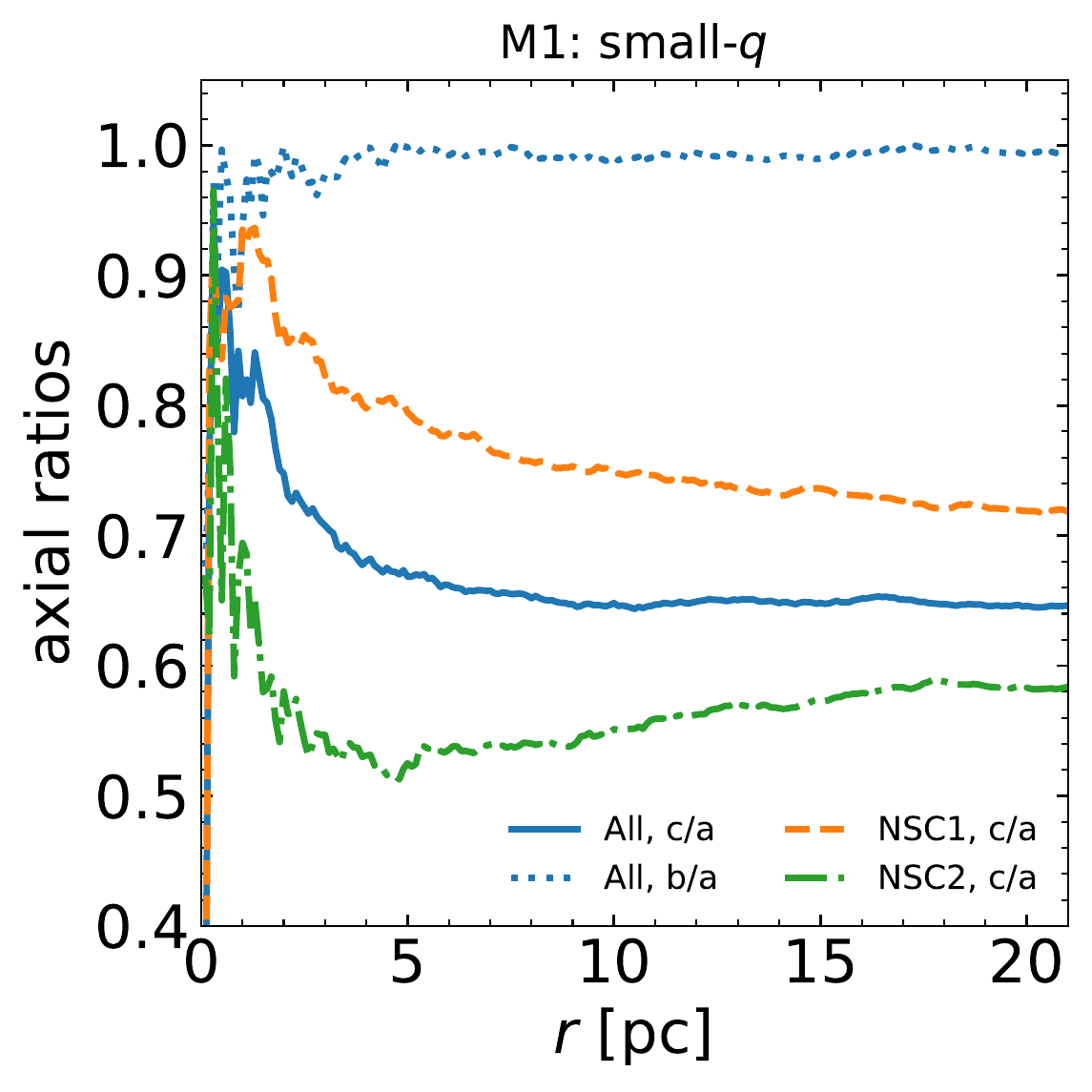}    \includegraphics[width=0.33\textwidth]{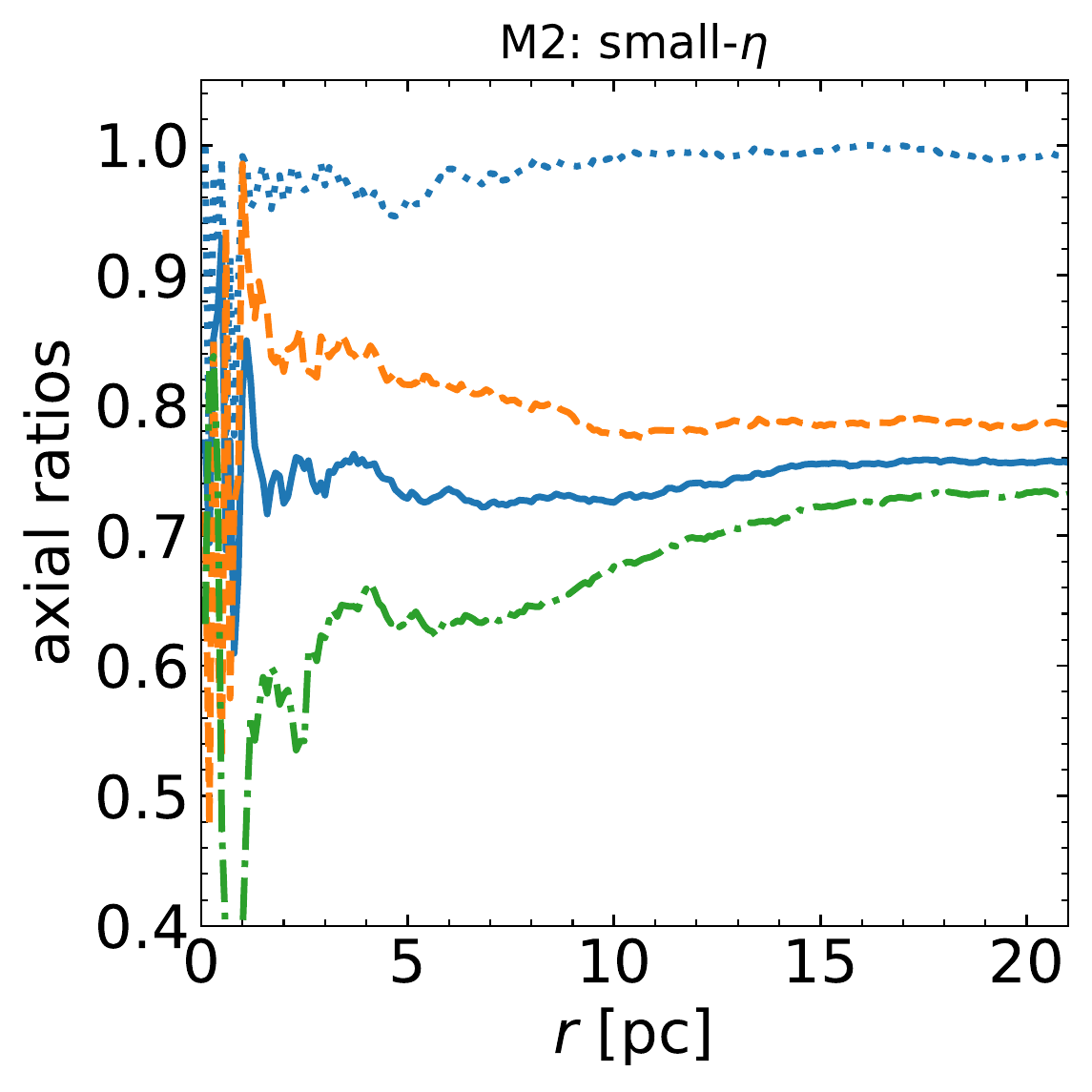}
    \includegraphics[width=0.33\textwidth]{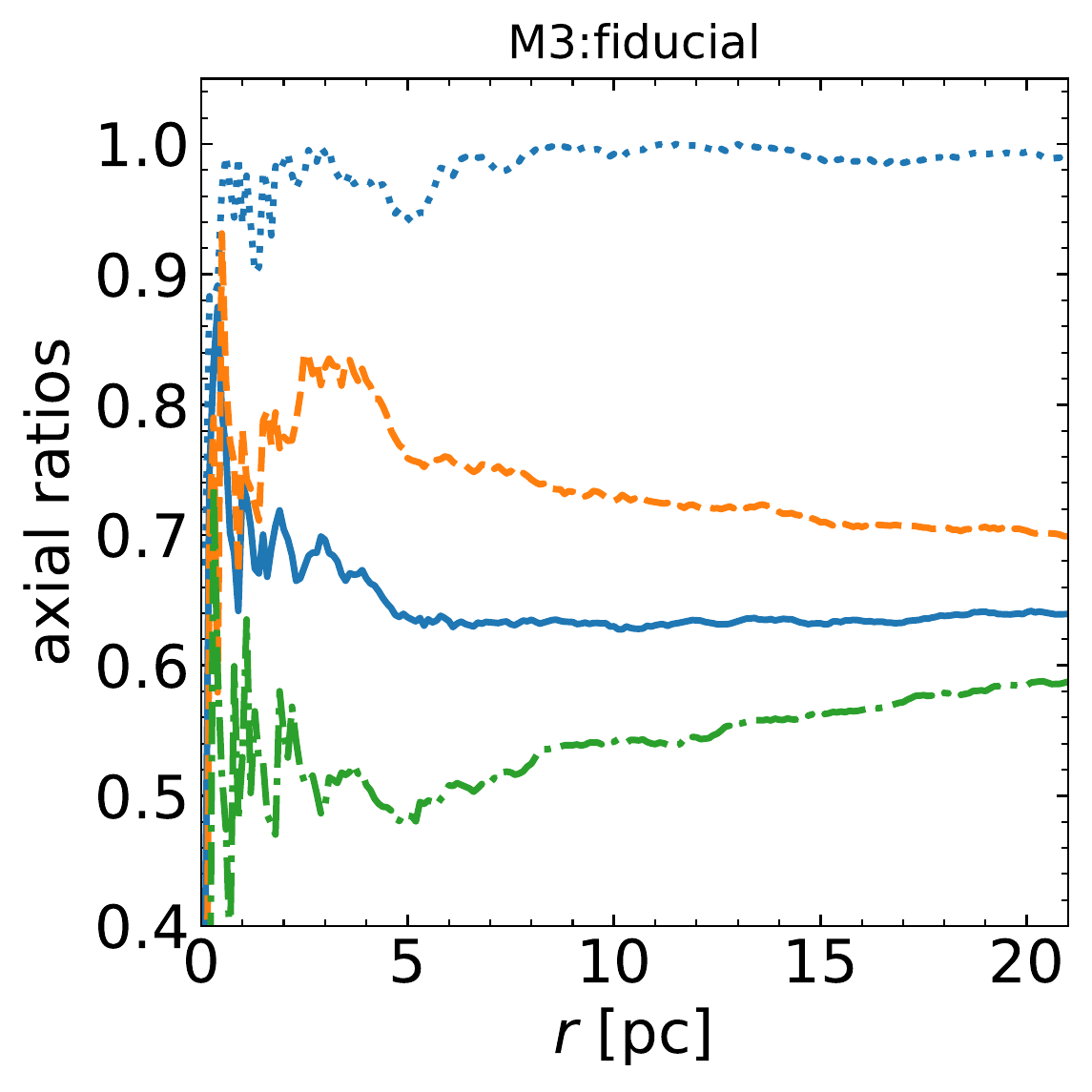}
    \includegraphics[width=0.33\textwidth]{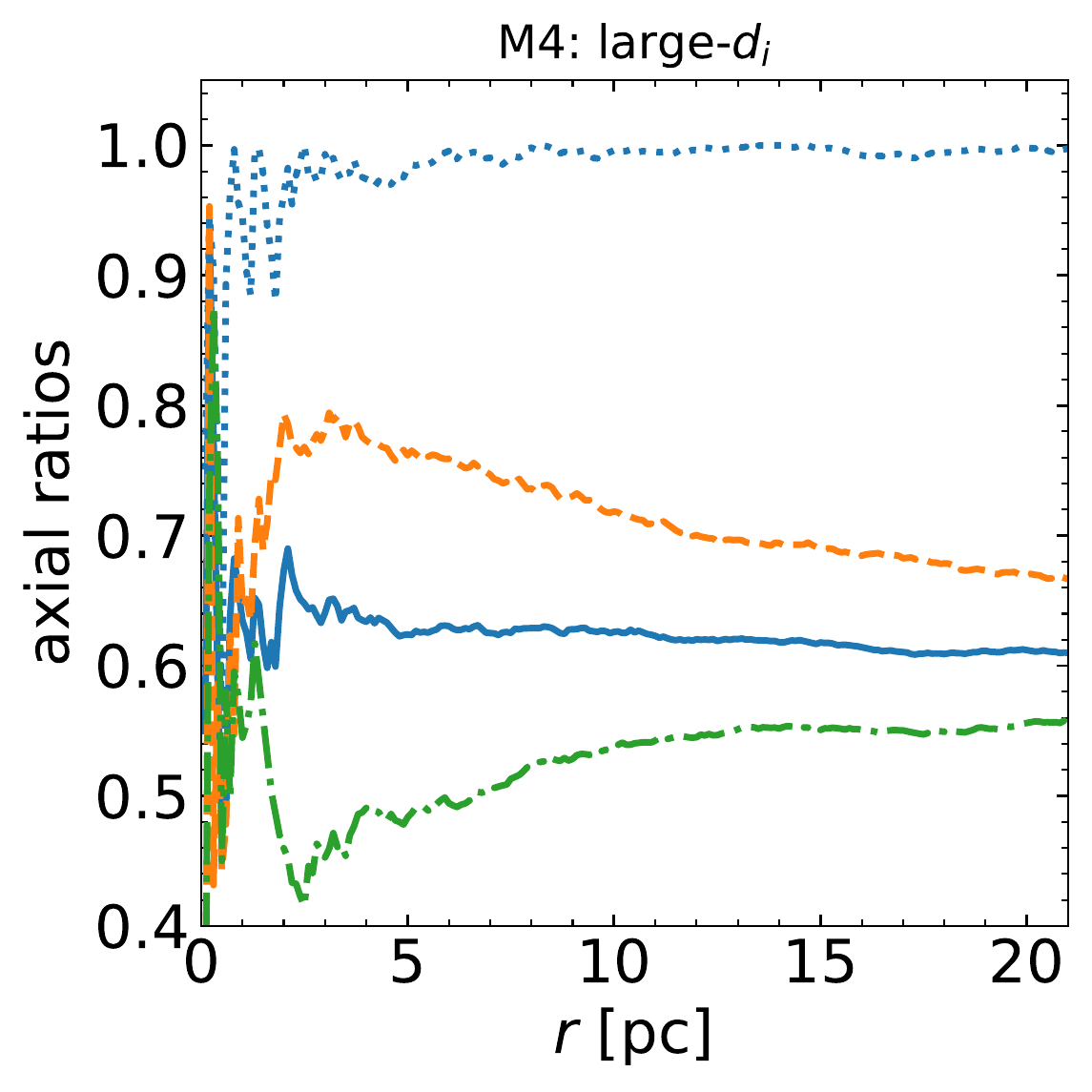}
    \includegraphics[width=0.33\textwidth]{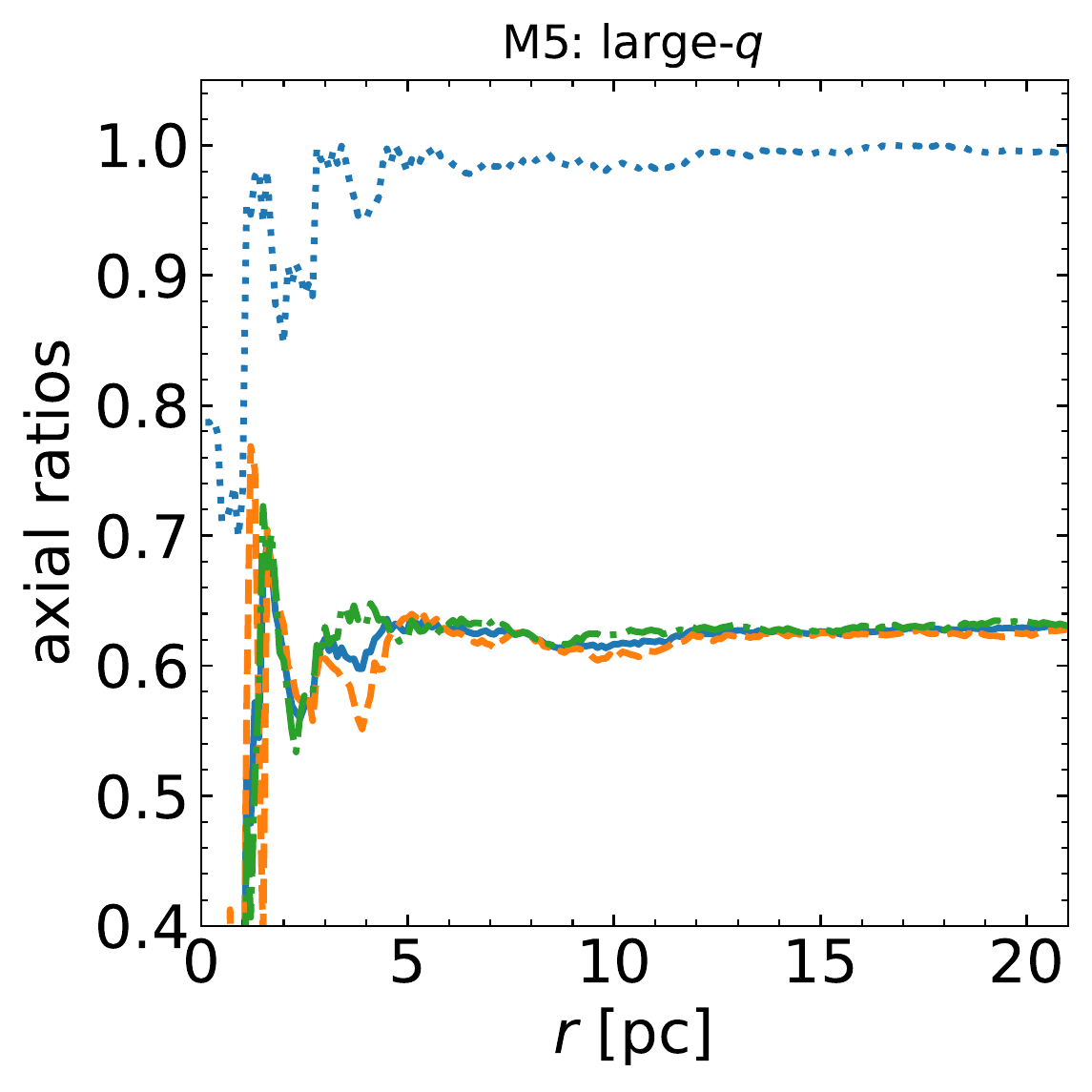}
    \includegraphics[width=0.33\textwidth]{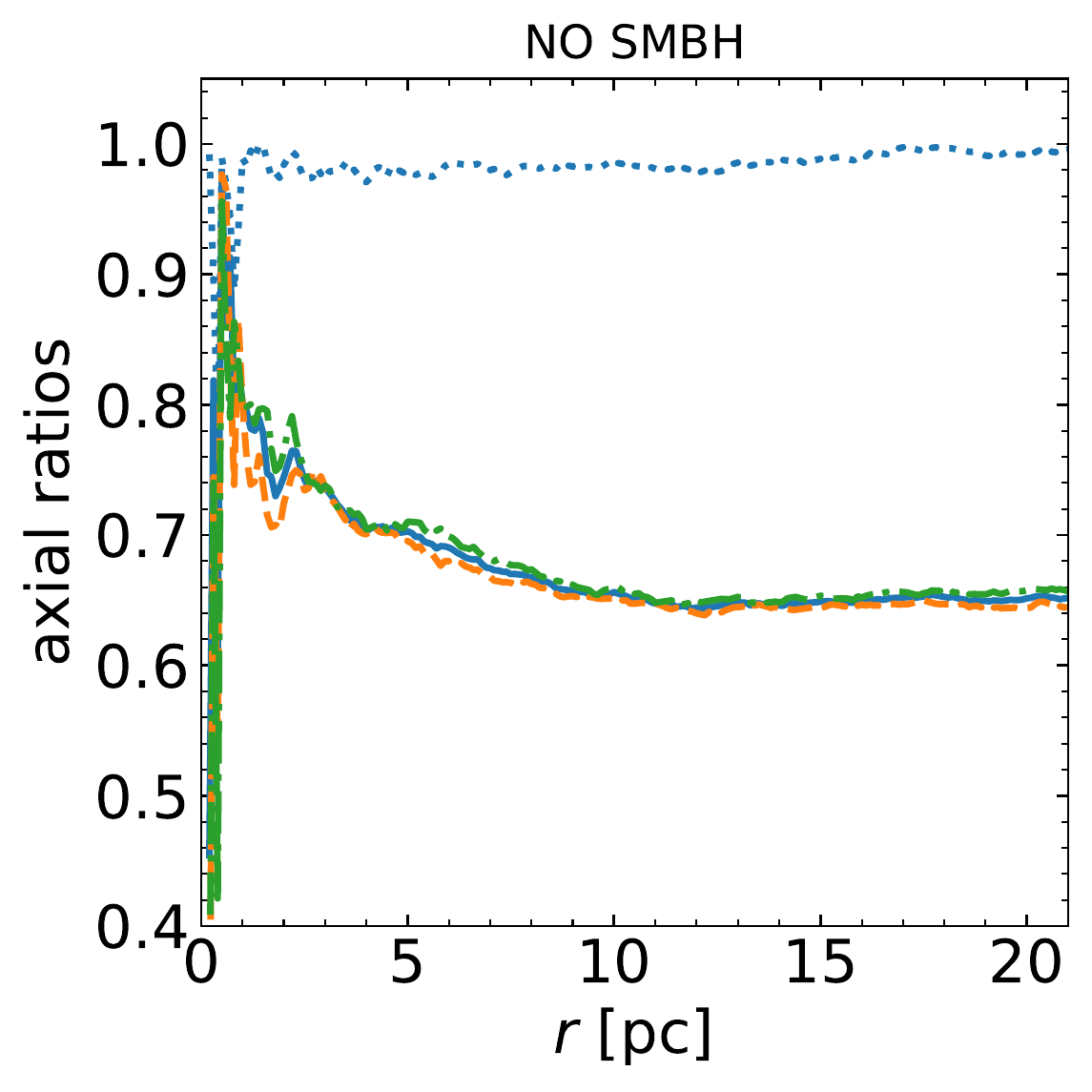}
    \includegraphics[width=0.33\textwidth]{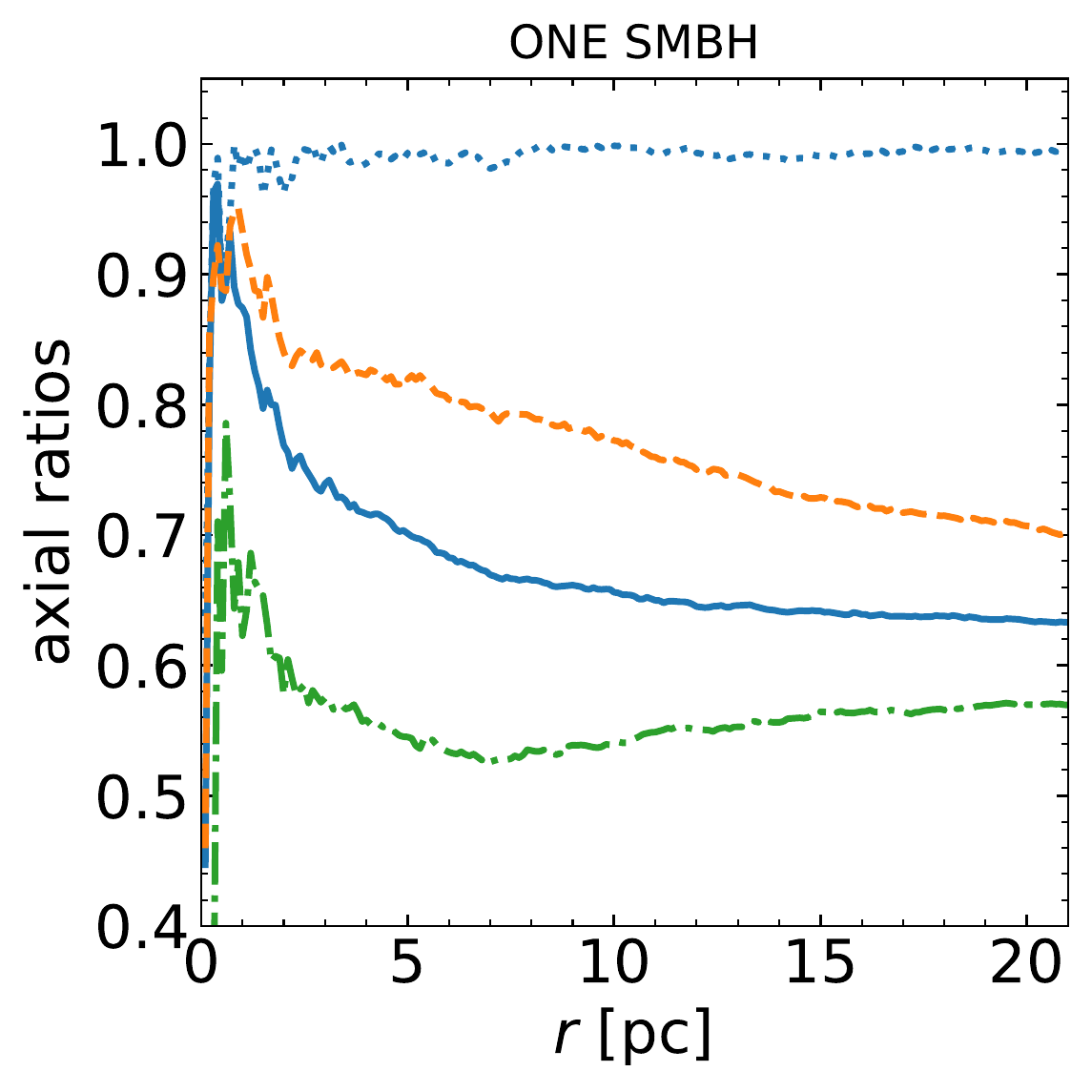}
    \caption{Axial ratios for the final NSCs obtained in our merger models. From top to bottom and from left to right the plots refer to model M1, M2, M3, M4, M5, NO SMBH and ONE SMBH models. The blue lines are for the axial ratios of the entire system. Both $b/a$ (dashed line) and $c/a$ (solid line) are plotted. The orange dashed line is for the $c/a$ axial ratio of NSC1 and the  green dot-dashed line is for the $c/a$ axial ratio of NSC2. The $b/a$ ratios for NSC1 and NSC2 are in all cases close to unity and are not shown in the plots.}
    \label{fig:ax_rat}
\end{figure*}

\section{Results} \label{sec:res}
%\go{Results from non-SMBH and one-SMBH runs should be merged into this section, including figs and tables.}
We analyse the final snapshot of all our merger simulations, which are taken at  time $t=20\,$Myr.   We identify the different models by the names listed in Table \ref{tab:tab1} for the models with two SMBHs and in Table \ref{tab:tab_app_comp} for those with no or only one SMBH.
All the NSC pairs with an SMBH each merge in a few Myr ($\sim 1-2$\, Myr) and, at the end of the simulation (i.e. after 20\,Myr), the SMBHB has formed and has significantly hardened. 
At this point, numerically following the evolution of the SMBHB becomes computationally expensive, significantly slowing down the calculation. To avoid this problem, \cite{Ogiya20} followed the evolution of the binary using an analytic approach, finding that the coalescence between the SMBHs is expected to take place between 57.6Myr and 5.3Gyr, depending on the initial conditions of the merger. The coalescence time is shorter for smaller SMBH mass ratios, while it is less dependent on the assumed orbital initial conditions.\\ In the NO SMBH and ONE SMBH cases, the merger takes 1-2 Myr longer  than in the two-SMBH cases (i.e. double the time with respect to those latter cases). The model with one SMBH only is the one that requires the longest merger time.  \\
%Since the SMBH coalescence affects the structure of the NSC only in the region enclosed within the sphere of influence of the BHs and leaves the external regions mostly unscathed, we neglect this process. 
We analyse the radial properties of the final NSC along its full extension to look for signatures left by the merger process and the formation and hardening of the SMBHB.
To avoid spurious effects due to escaping particles, in our analysis, we consider only particles bound to the system, i.e. particles with negative total energy with respect to centre of density of the system.  The unbound mass fraction is 2 per cent in the models with no SMBH or with only one SMBH, and it increases with the mass ratio between the SMBHs; it is 4 per cent in M1, 5 per cent in M2, 7 per cent in M3, 8 per cent in M4 and 12 per cent in M5. The unbound stars would be bound to the galaxy's potential if present. \\
Our tests are meant to understand the structural and dynamical properties of the final NSC imprinted by the merger process and the formation and hardening of the SMBHB that might help to identify NSCs that recently went through such a process and to find indirect observational hints of the presence of an SMBHB.
%%%
\subsection{The shape of the merger result}\label{sec:shape}
The NSC that forms after the merger is expected to be a flattened and rotating system, because of the conservation of the orbital angular momentum linked to the relative orbital motion of its progenitors. Our models have different amounts of initial orbital angular momentum, set through the parameters $\eta$ and $d_{i}$ (see Section \ref{sec:ICs} and Table \ref{tab:tab1}). 
Figure \ref{fig:map_xz} shows the surface density maps of our seven final NSCs, plotted considering an edge-on view, i.e., considering a line-of-sight perpendicular to the total angular momentum of the final NSC. 
 All the NSCs are significantly flattened and centrally concentrated. Figure \ref{fig:map_xz} visually shows that the M4 simulation is the most flattened NSC, while M2 is the least flattened one among our final NSC sample. Those are indeed the cases characterized by the largest and smallest initial orbital angular momentum values. %\textcolor{red}{The additional models run with no or only one SMBH also produce flattened NSCs. }}  \\
To quantify the flattening, we calculate the axial ratios of the final NSC following the approach described by \cite{Katz91}. This is an iterative method that uses the principal components of the inertia tensor to estimate the symmetry axes of the particles inside the spheroid of radius $r^2 = x^2/a^2 + y^2/b^2 + z^2/c^2$; we set a precision of $5\times10^{-4}$ as the convergence criterion.
In the definition of the radius of the spheroid, $a$, $b$  and $c$ are the major, intermediate and minor axis of the ellipsoid, respectively.
The axial ratios of the final NSCs and of the particles initially belonging to their two progenitors are shown in Figure \ref{fig:ax_rat}.
As the intermediate-to-major axial ratio, $b/a$, is approximately equal to unity at any radius, all merged NSCs are oblate\footnote{The intermediate-to-major axial ratio, $b/a$, is approximately equal to unity also for the final NSC1 and NSC2, in each of the runs.}. In the two-SMBH models, the stars that initially were in the NSC hosting the least massive central SMBH (NSC2) are always in a more flattened 
%\go{Why? Because the stars in NSC2 are dynamically colder than counterparts in NSC1?} 
configuration compared to the stars initially in the NSC hosting the most massive SMBH (NSC1). 
Dynamical friction, as well as the ouroboros effect, are less effective on the secondary NSC and on its SMBH. NSC2, therefore, retains a larger amount of its initial orbital angular momentum, leading to its larger observed flattening. At the same time, the stars in the primary NSC are dynamically heated more efficiently by the more massive SMBH, keeping them in a more spherical configuration.
The difference between the flattening of the two stellar populations is significant in the case of the simulations with $q=0.01$ (M1) and $q=0.1$ with $\eta=1.0$ (M3, M4), and does not depend on the initial distance between the NSCs. In the model with $q=0.1$ and $\eta=0.5$ (M2) the two populations forming the final NSC have similar flattening at radii larger than $10\,$pc. The radius of transition corresponds roughly to the half-mass radius of the NSC (see Table \ref{tab:tab1}). In the case of the merger between two NSCs hosting SMBHs of the same mass (M5), the two progenitors merge and both attain a flattening approximately equal to $0.6$. In all cases, the flattening of the final NSC at the half-mass radius is between $0.6$ and $0.7$, a value similar to the one observed for the Galactic NSC \citep{Schoedel14a}. 
 In the NO SMBH case, similarly to model M5, we have that $c/a$ is the same for the two populations. However, while for M5 the cluster is centrally flattened with $c/a<0.7$, the NO SMBHs final NSC is centrally spherical and has $c/a>0.7$ at NSC-centric distances smaller than 5\,pc. This is a consequence of the longer decay time expected in the absence of the SMBHs, that allows the systems to have more time to centrally relax and settle on an internal spherical configuration. In addition, the absence of a central perturber let the system dynamically relax in a more efficient way. %\go{It might be better rephrase it a bit as no particles practically feel dynamical friction in NO SMBH run. I guess the center dynamically relaxes well in the absence of the perturber, i.e. SMBH?} 
When only one SMBH is present, the $c/a$ ratio for NSC1 shows a behaviour and values similar to what seen for model M1, while NSC2 is centrally more flattened than in M1. The entire NSC is more centrally spherical than the NSC obtained in M1. A larger amount of angular momentum is retained by NSC2 when it does not contain an SMBH (as for NSCs that host smaller SMBHs), causing it to be more flattened than in M1. 
%\go{Why? Good to discuss the physical origins of these structures.}
\\
\subsection{The density profile and cumulative mass of the merger result and of its components}
\begin{figure*}
    \raggedright
    \includegraphics[width=0.33\textwidth]{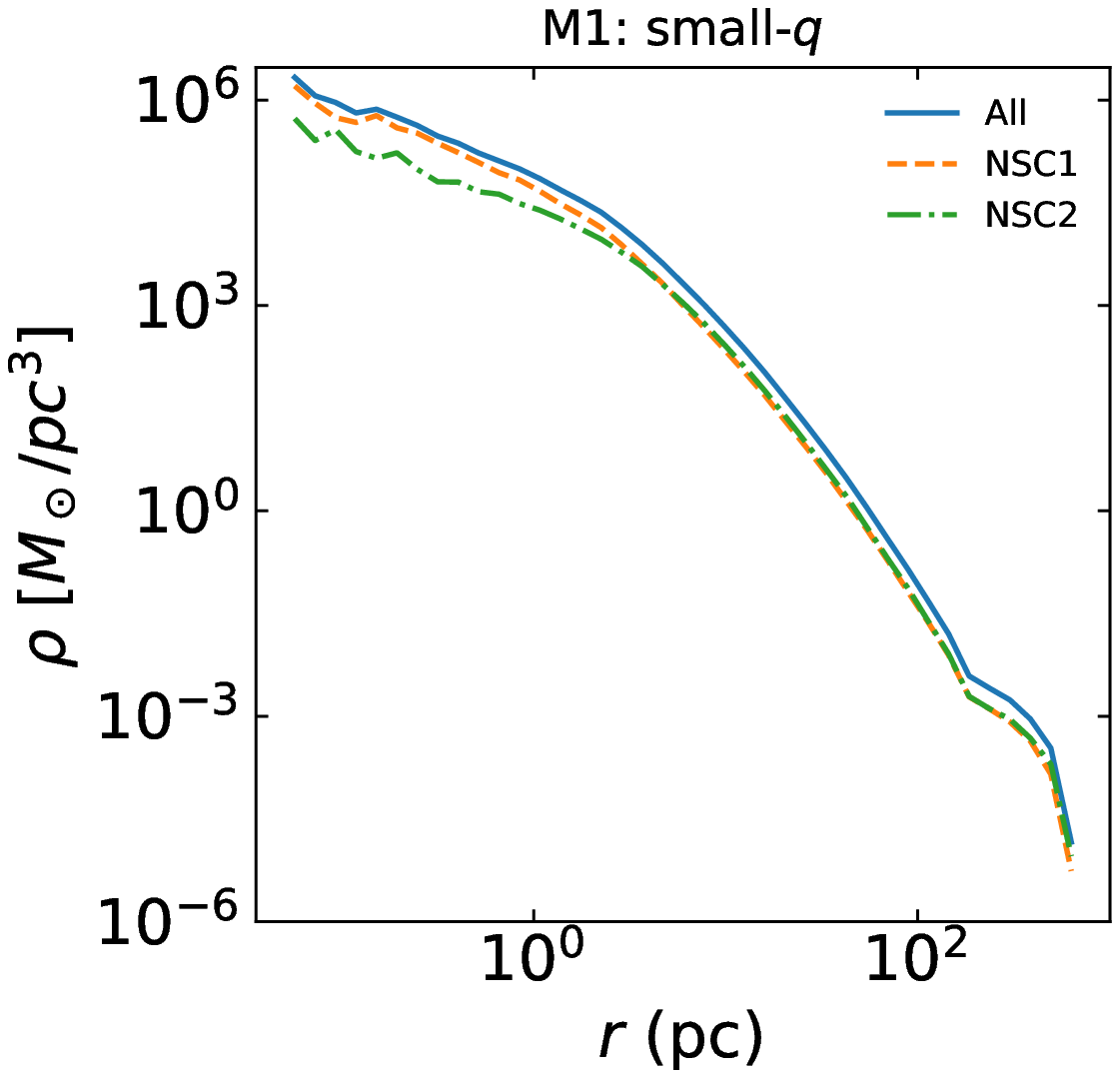}
    \includegraphics[width=0.33\textwidth]{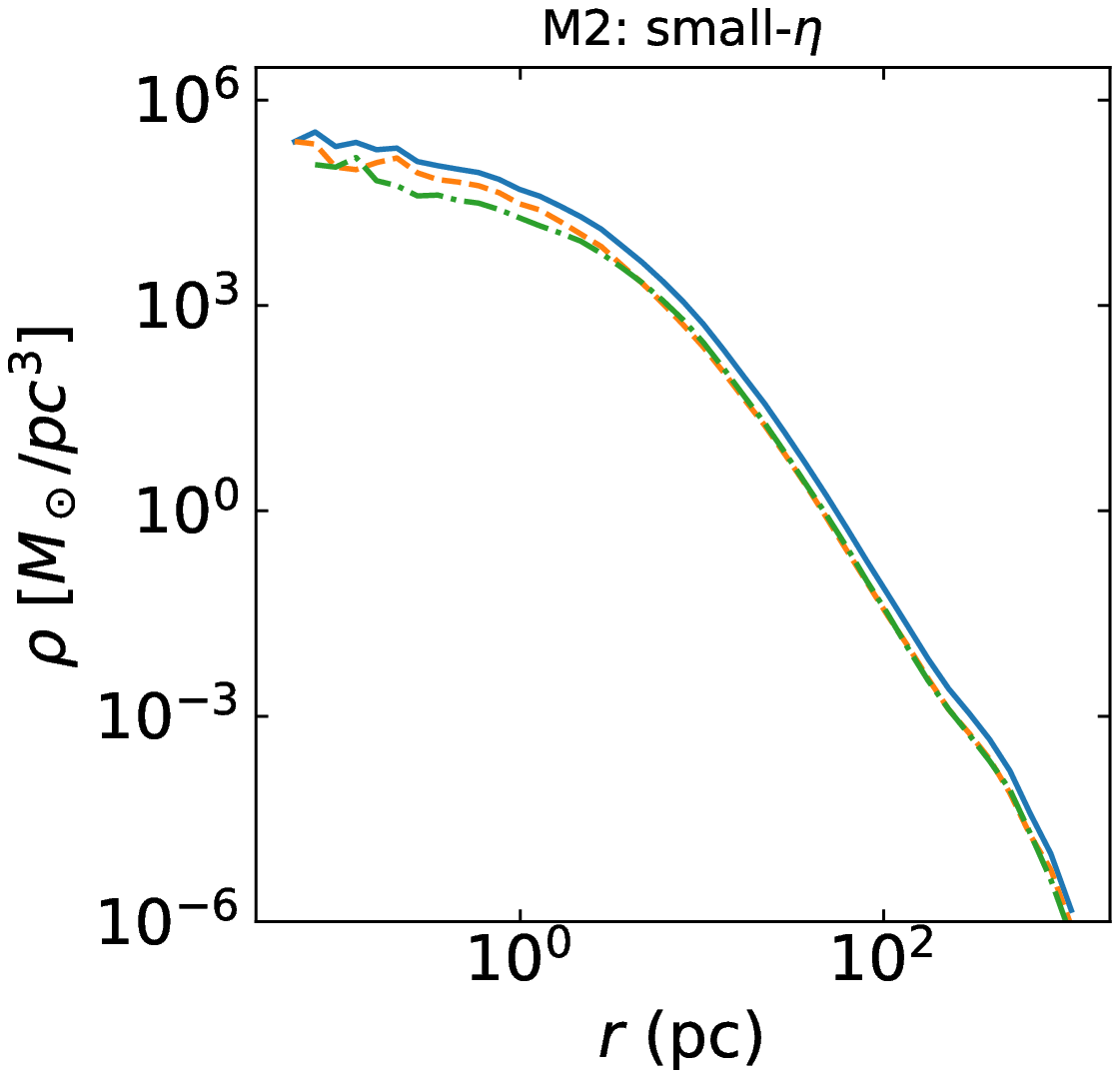}
    \includegraphics[width=0.33\textwidth]{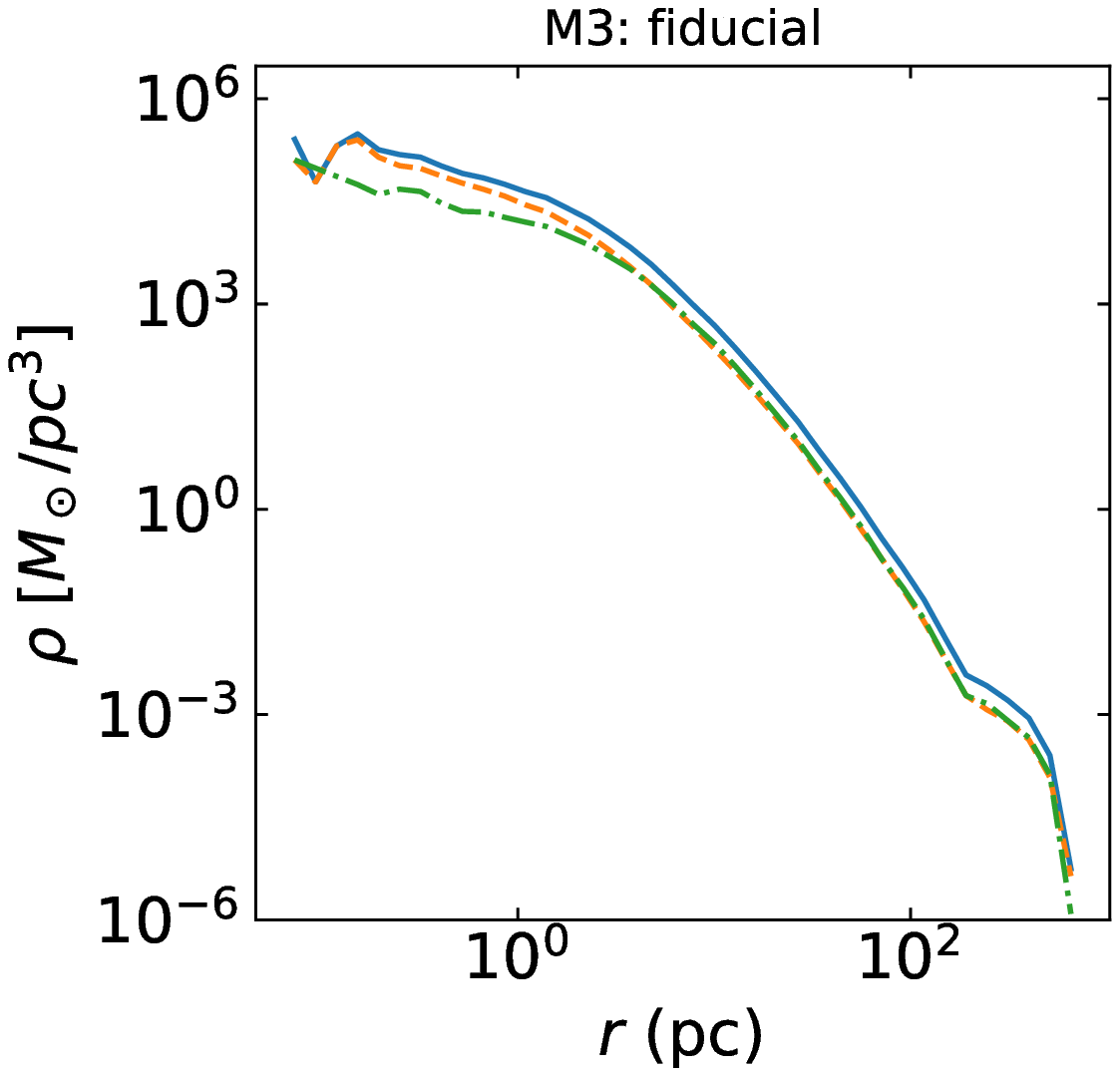}
    \includegraphics[width=0.33\textwidth]{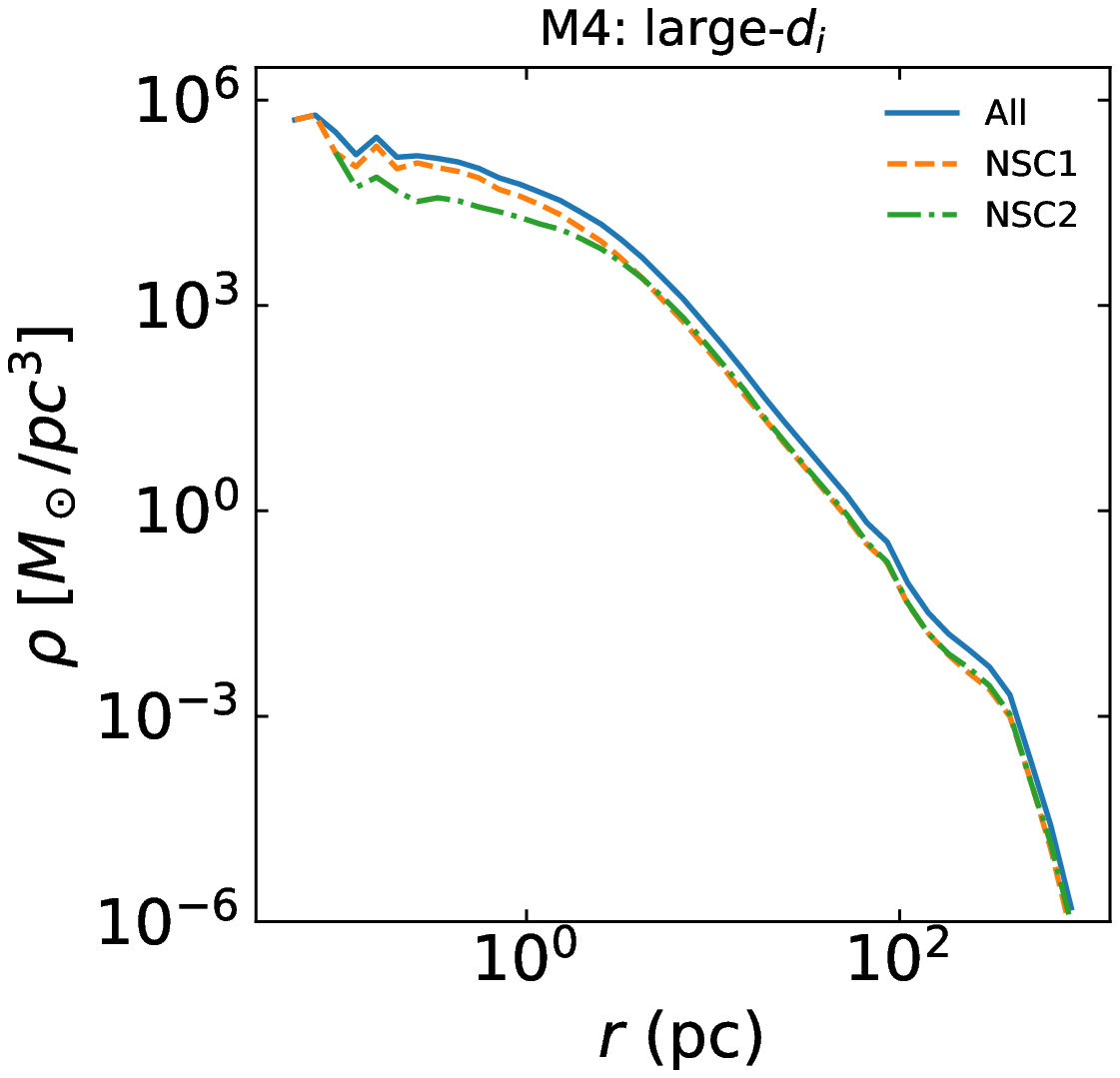}
    \includegraphics[width=0.33\textwidth]{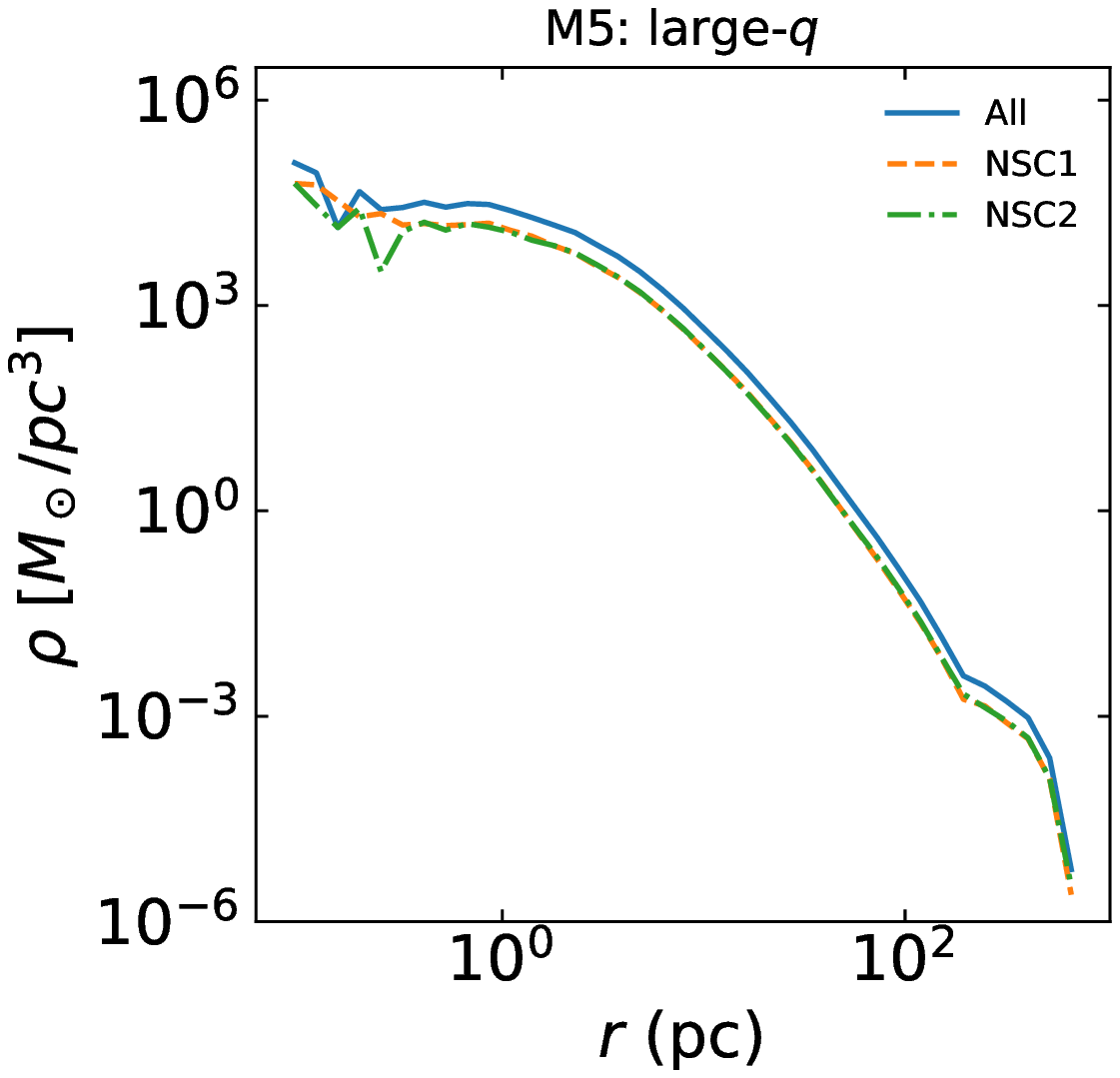}
    \includegraphics[width=0.33\textwidth, trim= 0cm 0 0 0cm , clip]{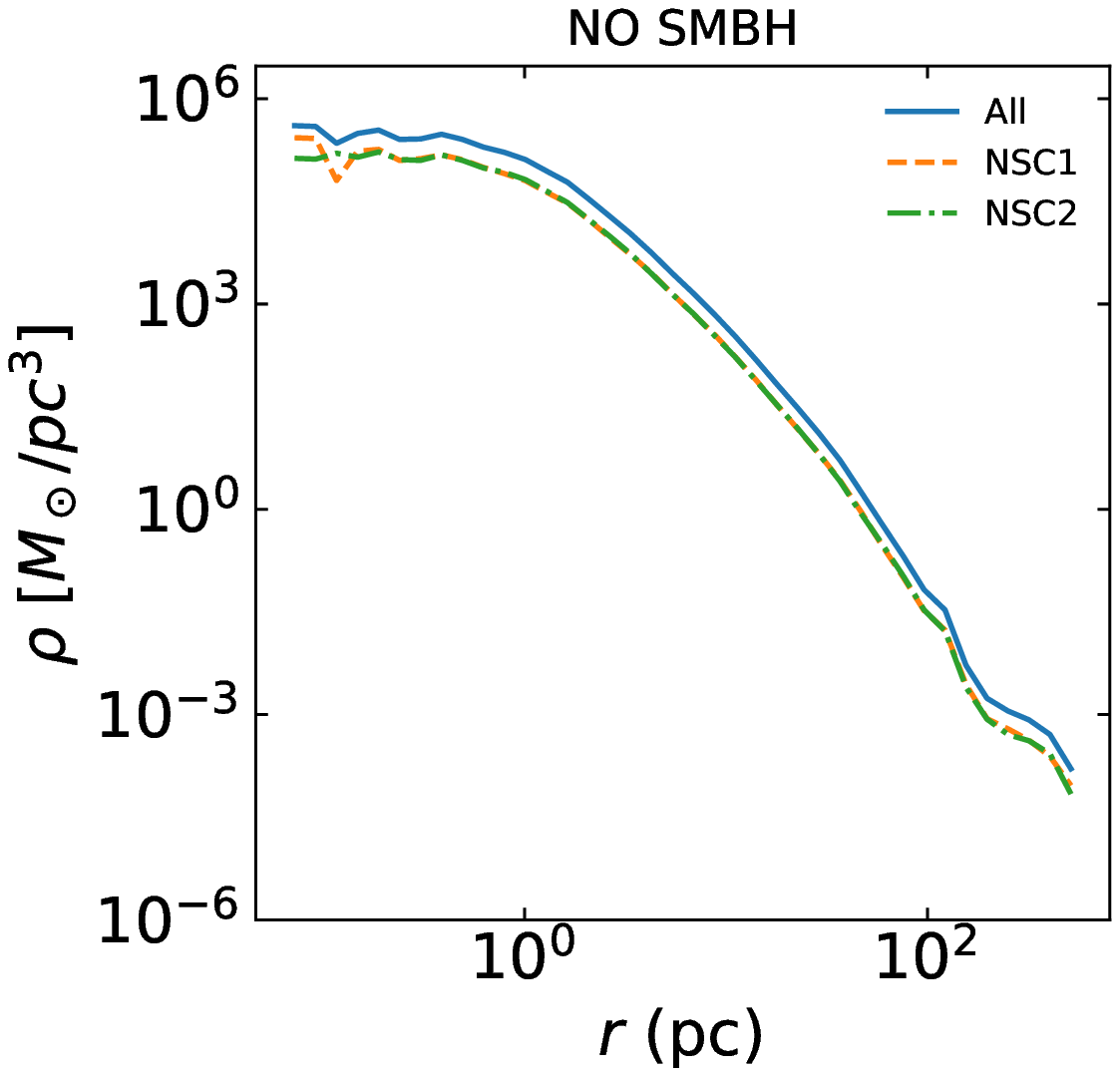}
    \includegraphics[width=0.33\textwidth, trim= 0cm 0 0 0cm  , clip]{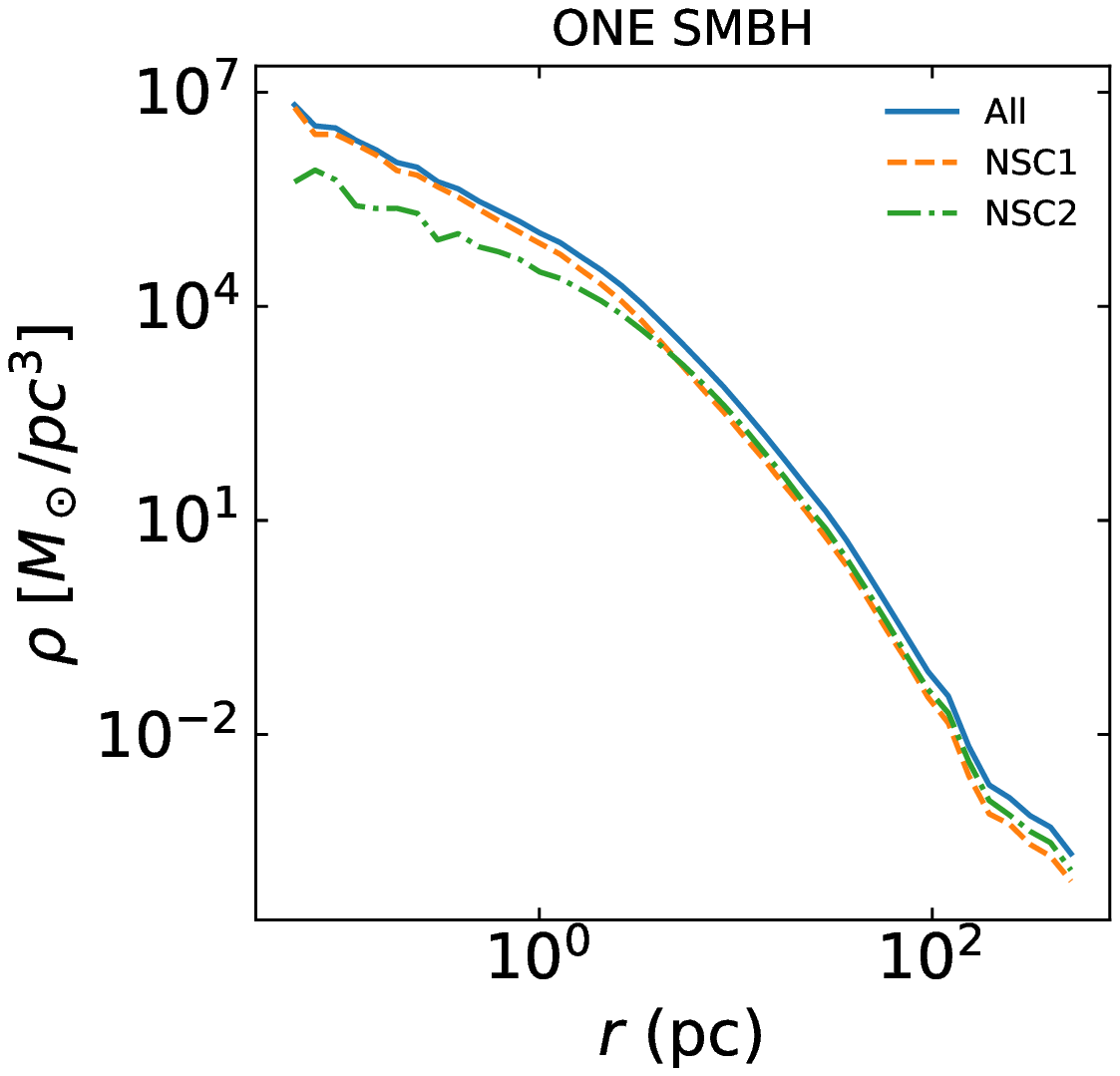}
    \includegraphics[width=0.33\textwidth, trim= 0cm 0 0 0cm  , clip]{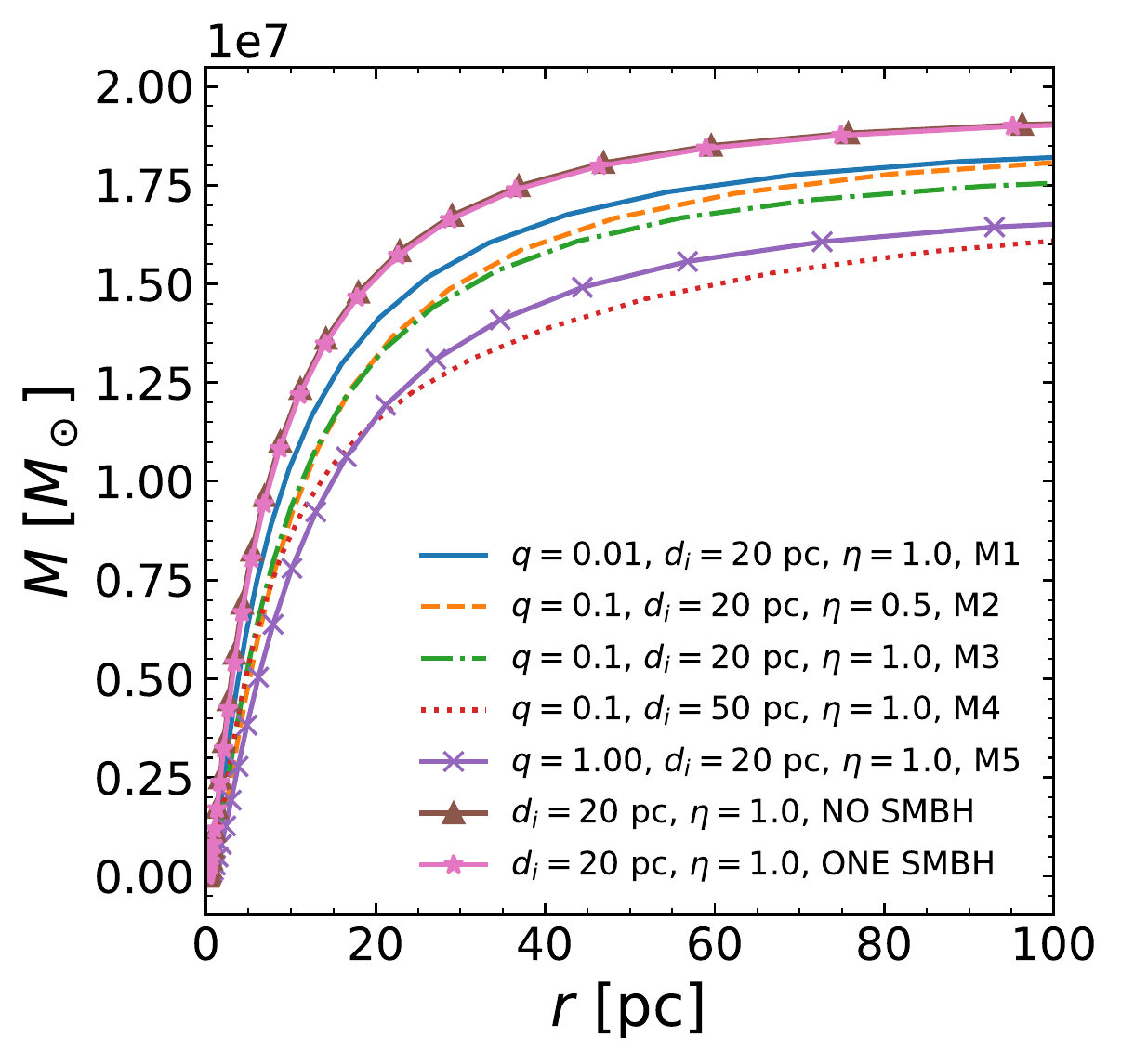}
    \caption{
    Spatial density profiles for the final NSC in our models. From top to bottom and from left to right the plots refer to model M1, M2, M3, M4, M5, NO SMBH and ONE SMBH models. The density profile of the entire system is shown using a solid blue line, the density profile of the stars initially belonging to NSC1 and NSC2 are shown using an orange dashed line and a green dot-dashed line, respectively. The bottom right panel shows the cumulative mass of the systems as a function of the NSC-centric distance.}
    \label{fig:density}
\end{figure*}

The central density of the NSC that forms after the merger varies significantly with the initial merger conditions (see Figure \ref{fig:density}). 
Table \ref{tab:tab1} shows the half-mass radii of all the final NSCs and their total masses along with the mass enclosed within $d_i$. The half-mass radius increases with the mass ratio between the SMBHs ($q$), while the total mass shows a slight decrease with $q$. Each of the two stellar populations contribute to approximately half of the final NSC mass enclosed within $d_i$.
The central density of the M1 final NSC is larger than $10^6\,\Mo$/pc$^3$. This is the largest value observed for our two-SMBH models. All the other cases show a central density smaller than $\sim10^6\,\Mo$/pc$^3$. 
The population coming from NSC2 is always less centrally concentrated than the one brought in by NSC1, as it was in the progenitor NSC. 
A small amount of stars gravitationally bound to the NSC are observed as far as 1\,kpc from the NSC centre. These stars are scattered at those large distances during the merger. We should, however, caution that no underlying galactic potential is  considered in the simulations. In a more complete set up, a fraction of these stars scattered at distances larger than few times the half-mass radius from the final NSC centre will be captured by the tidal field of the host galaxy. These stars will no longer be bound to the NSC, and become part of other central structures of the galaxy (e.g. the bulge). Part of these stars could also contribute to the total mass of the nuclear stellar disc of the galaxy (see Section \ref{sec:nsd}).\\
While the M1 NSC shows a steep central cusp in the density profile, the other systems are characterized by shallower density profiles (M2, M3, M4) or by a core-like central profile (M5). \\
The M4 cluster, whose progenitors start at a distance of 50\,pc, has  a smaller total mass compared to the M3 NSC which has similar initial conditions except for the smaller initial distance between the progenitors. This is due to the fact that a larger initial distance corresponds to a longer decay time and, therefore, to a larger mass loss, leading to a final NSC with a smaller mass. 
The M5 NSC is the least massive system among the ones formed starting from NSCs at an initial distance of $20\,$pc. This NSC is, as well, the least centrally concentrated final NSC observed in our sample. 
 When no SMBH is present the two populations coming from NSC1 and NSC2 follow the same density profile, as in M5, which is the two-SMBH model in which we adopted an SMBH mass ratio $q=1.0$. However, the NO SMBH model shows a smaller central core compared to the M5 NSC. When only one SMBH is present, the population coming from NSC1, which initially hosts the SMBH, forms a central steep cusp in the density profile due to gravitational contraction by the SMBH, reaching a central density one order of magnitude higher than what observed for the stars that were initially in NSC2.\\
The cumulative mass of the NSCs has a different radial behaviour depending on the merger conditions. This quantity seems to increase with decreasing $q$. The total mass of each final NSC and the mass of the same NSC calculated within $d_i$ are listed in Table \ref{tab:tab1}. The NSC that forms in M1 is the most centrally concentrated and massive system in our sample. The initial amount of angular momentum does not change significantly the mass accumulated within 40\,pc, while outside this radius the system with the lowest amount of initial angular momentum (M2) seems to be able to retain a larger fraction of the initial stars. The M4 NSC shows a significantly different cumulative mass distribution compared to the analogue simulation with smaller initial distance between the two SMBHs (M3). This departure is observed at an NSC-centric distance approximately equal to the half-mass radius of the system, and the difference is due to the fact that, given the larger initial distance, stars in M4 are distributed on a larger volume than in M3 during the merger. The NSCs obtained in the NO SMBH and ONE SMBH cases have a larger final bound mass -- approximately equal to the initial total stellar mass -- compared to the models run with two SMBHs -- and are more concentrated (see last panel of Figure \ref{fig:density} and Table \ref{tab:tab_app_comp}.  These results suggest that there are two effects to determine the final density profile and mass of the merger remnant; scattering to reduce the central density and contraction to increase it. The final density profile would be determined by the balance between them. Our results are explained by the fact that, in the absence of one or two SMBHs, as well as for low values of $q$, contraction dominates, as scattering due to the presence of an SMBHB is not in action, leading to higher central densities and cumulative masses. 

\begin{figure*}
    \raggedright
    \includegraphics[width=0.33\textwidth]{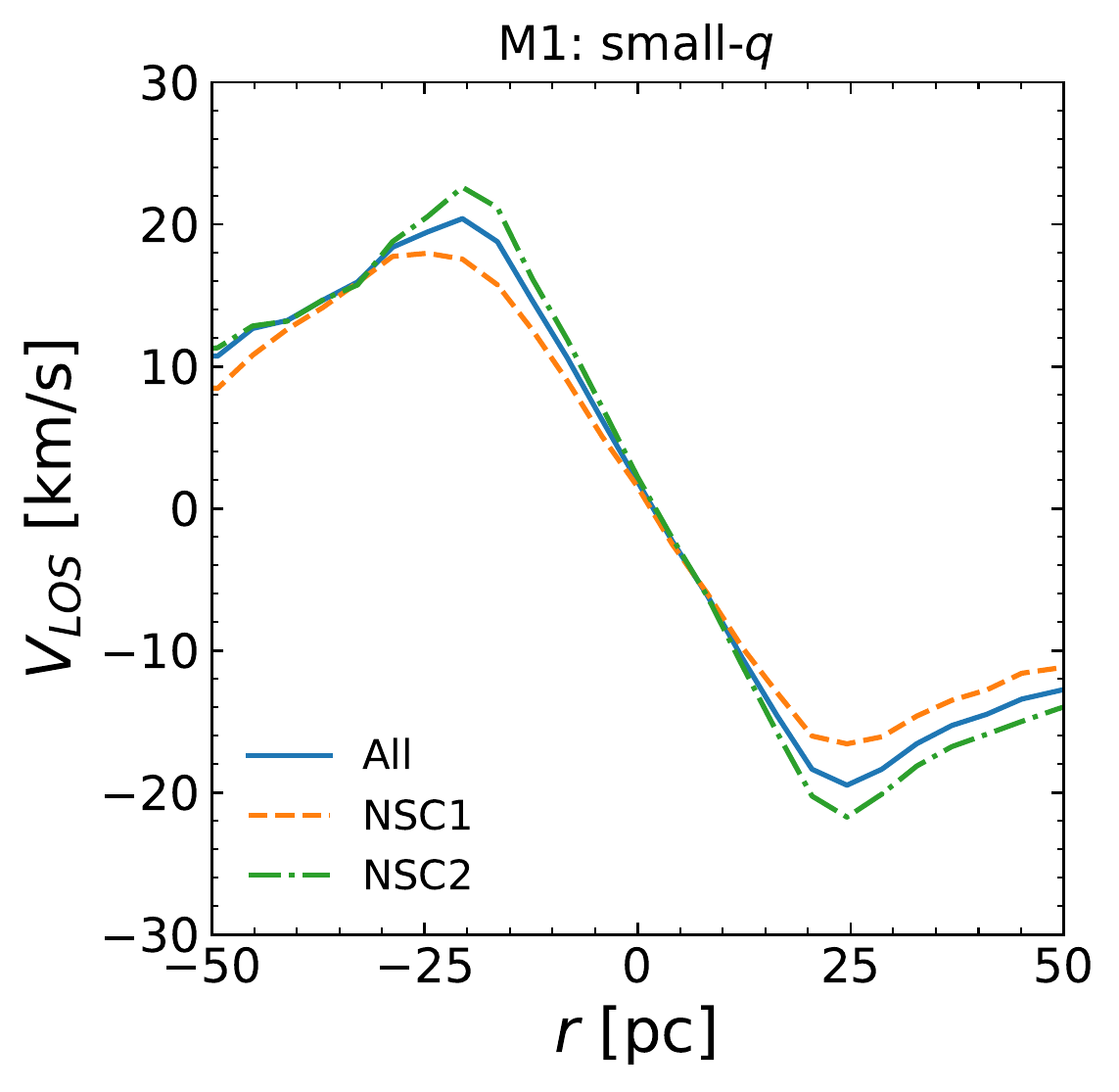}
    \includegraphics[width=0.33\textwidth]{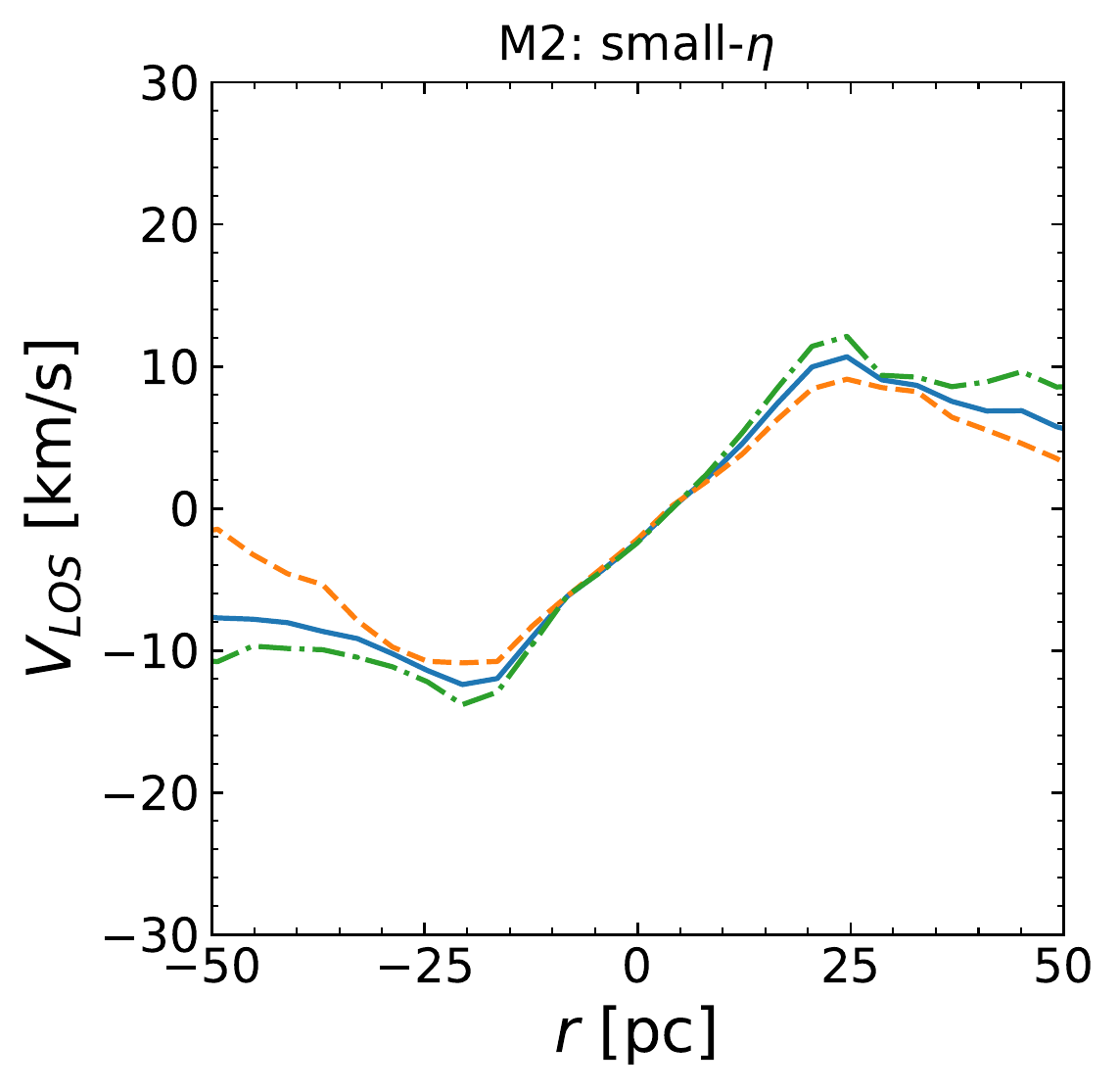}
    \includegraphics[width=0.33\textwidth]{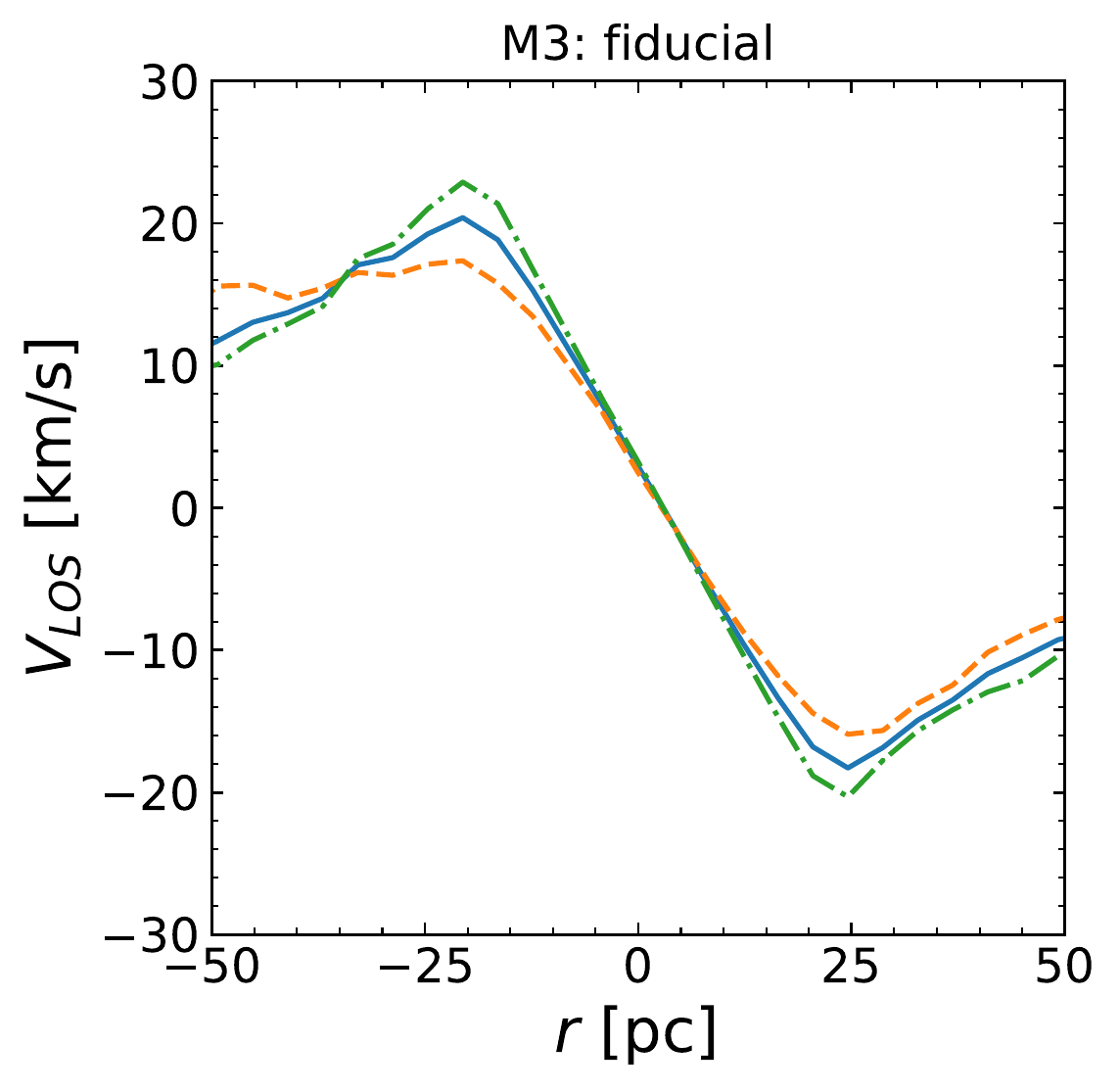}
    \includegraphics[width=0.33\textwidth]{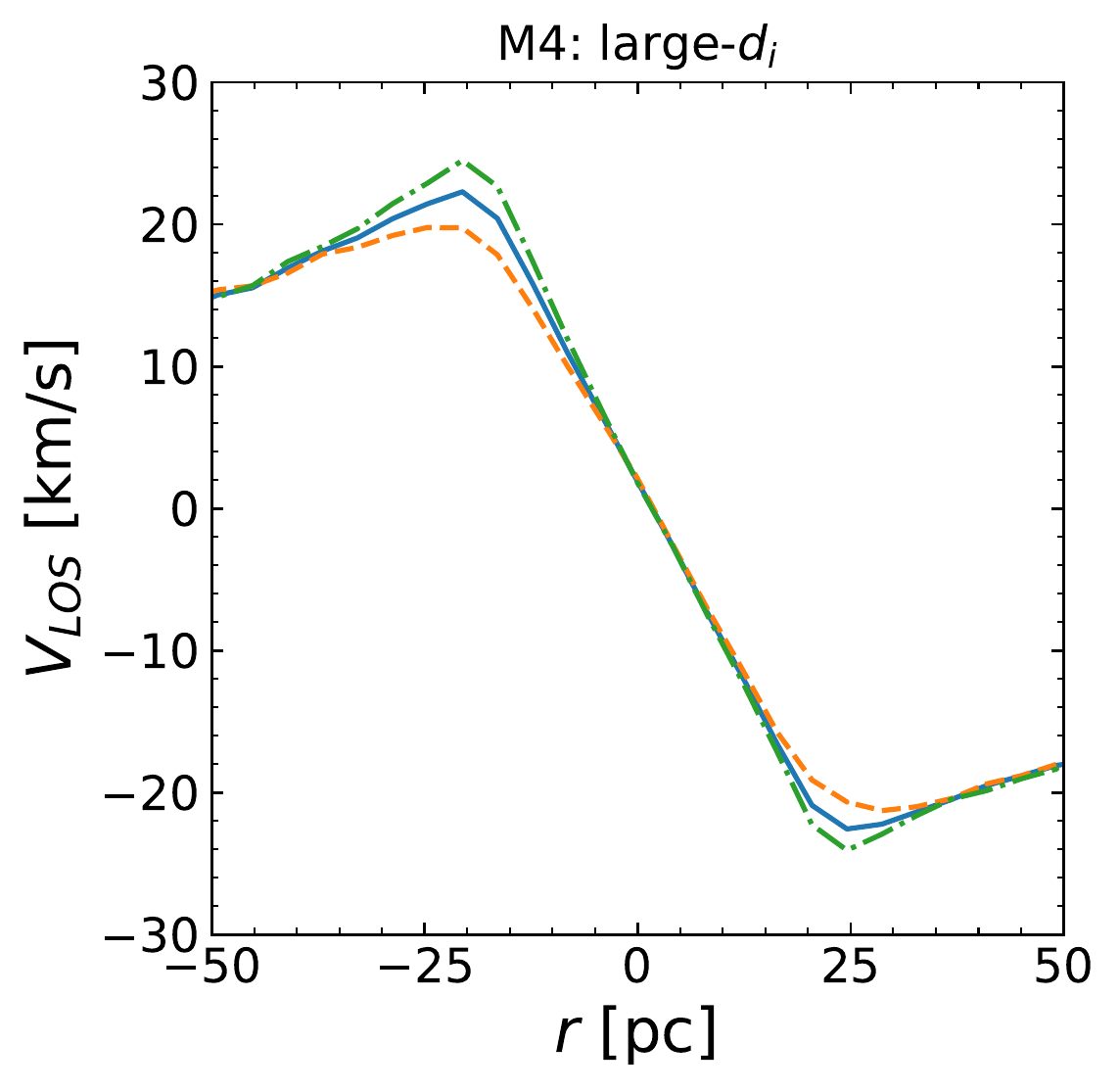}
    \includegraphics[width=0.33\textwidth]{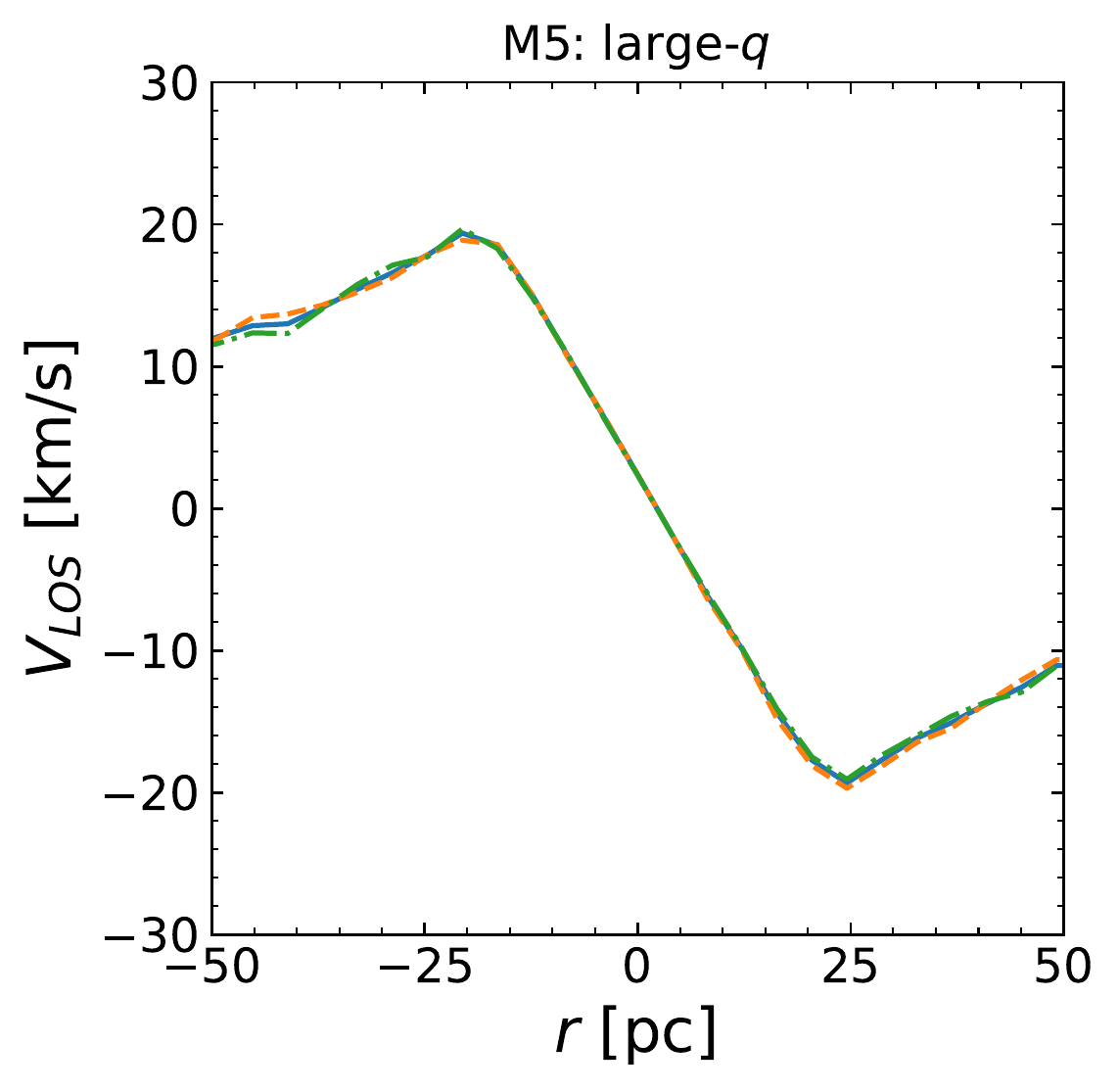}
    \includegraphics[width=0.33\textwidth, trim= 0cm 0 0 0cm, clip]{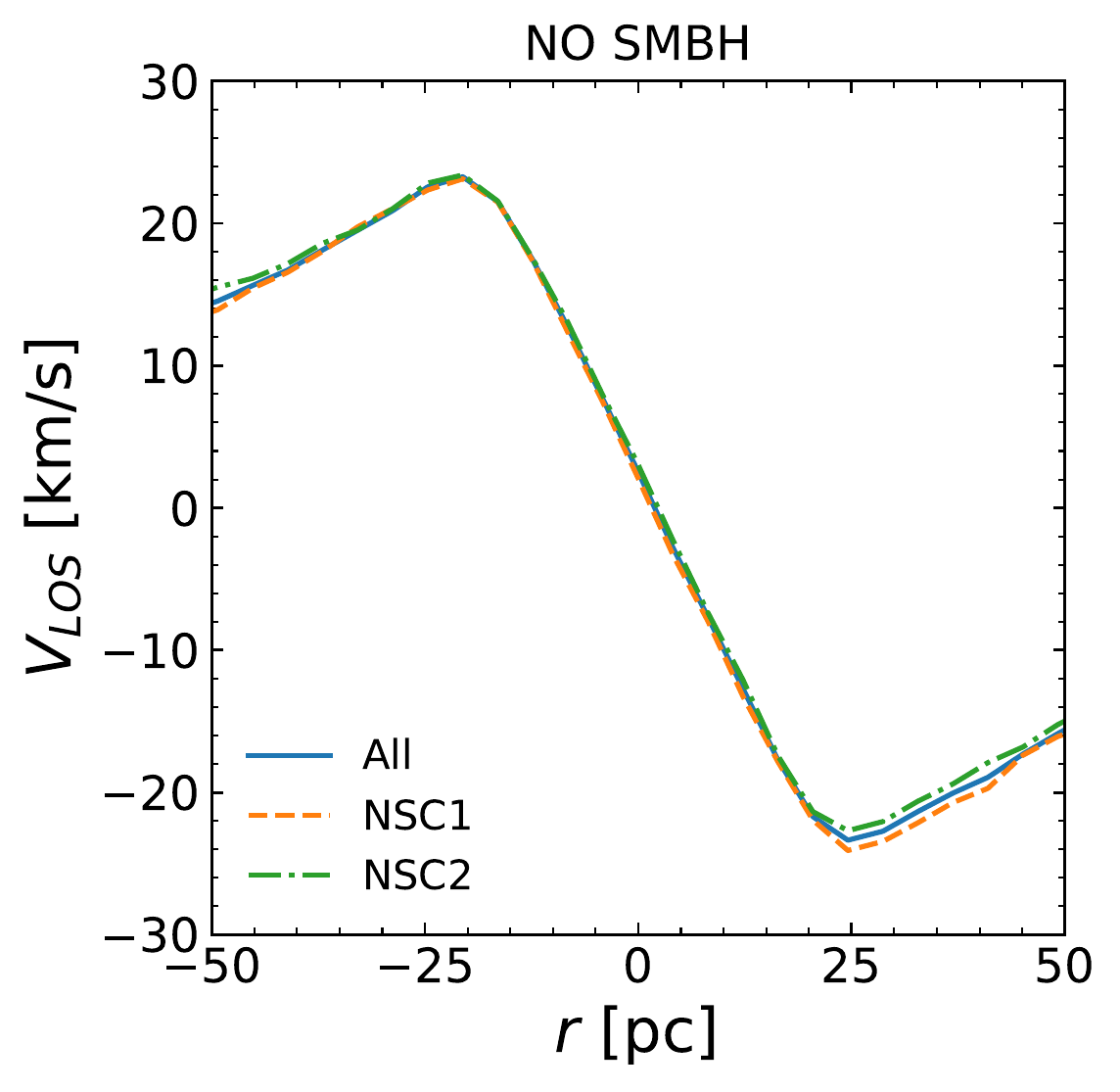}
    \includegraphics[width=0.33\textwidth, trim= 0cm 0 0 0cm, clip]{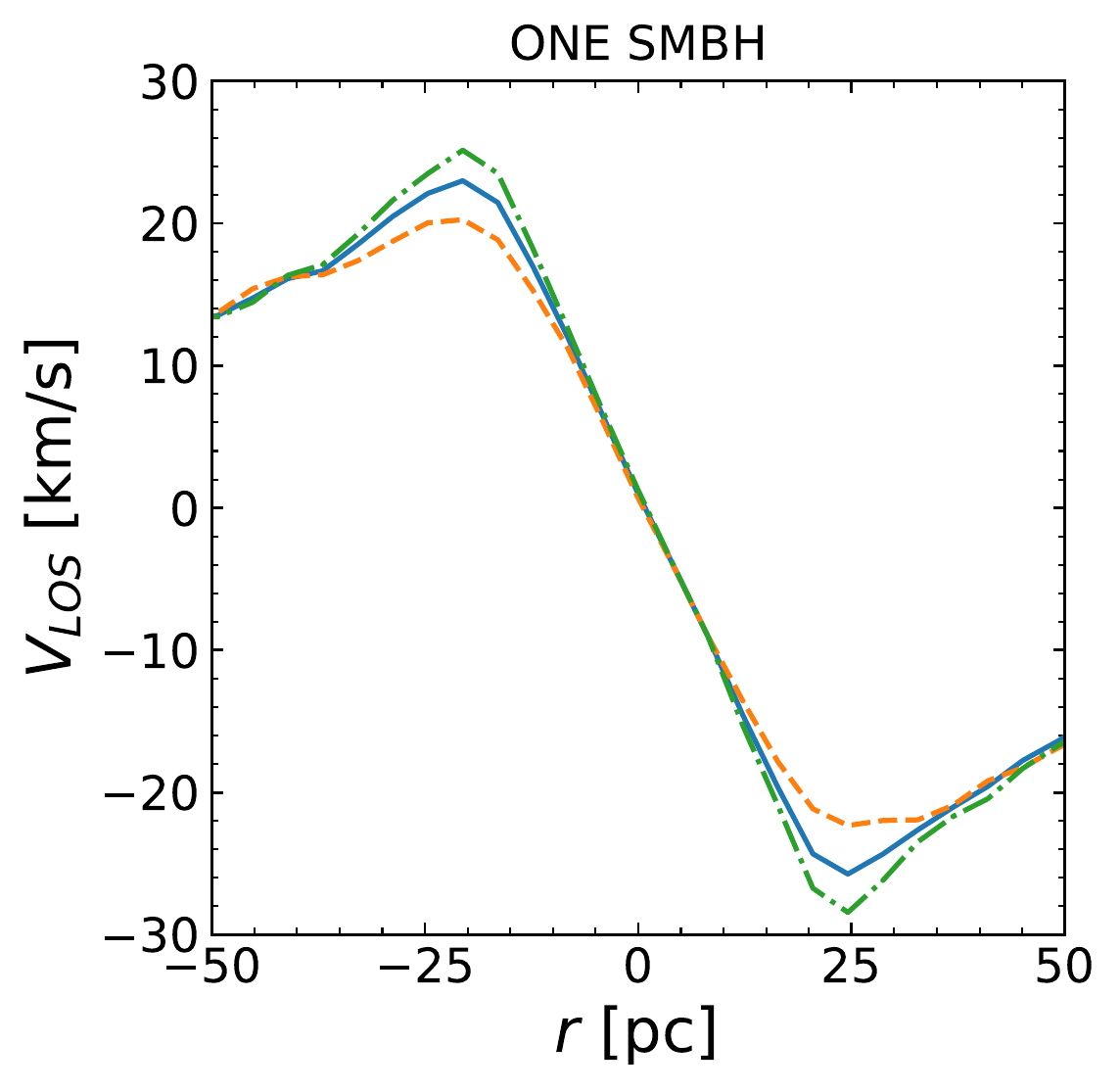}
    \caption{Rotation curves for our models (M1 to ONE SMBH going from top to bottom and from left o right). The solid blue line is for the rotational velocity of the entire final NSC. The  orange dashed line is for the stars initially in NSC1 and the green dot-dashed line is for the stars initially in NSC2. All the systems rotate, however, the amount of rotation and the differences between the rotational pattern of the two components depend on the initial conditions assumed in each model.}
    \label{fig:vel_curve}
\end{figure*}

\begin{figure*}
    \raggedright
     \includegraphics[width=0.45\textwidth, trim= 0cm 0 0 0cm, clip]{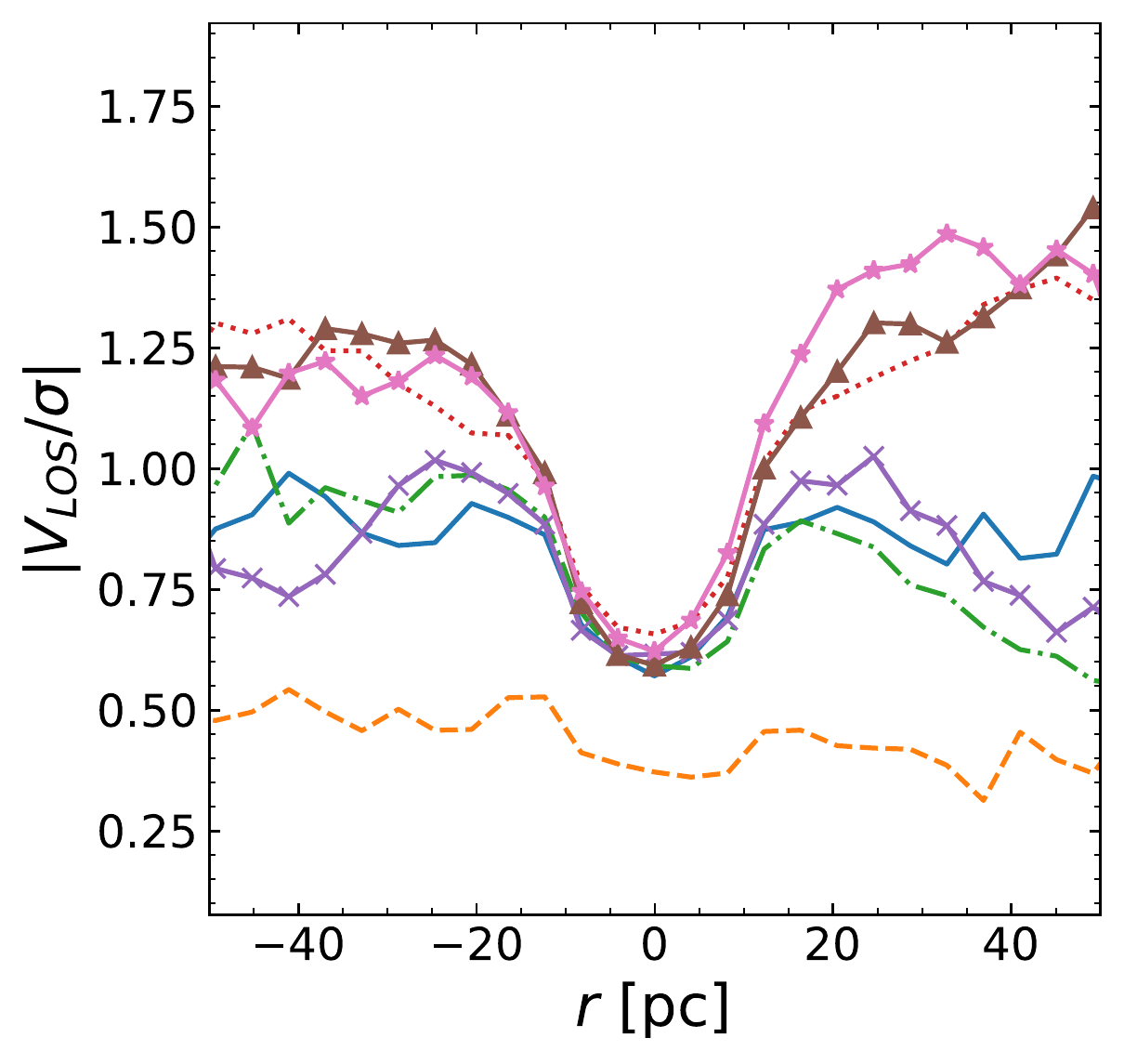}
     \includegraphics[width=0.465\textwidth, trim= 0cm 0 0 0cm, clip]{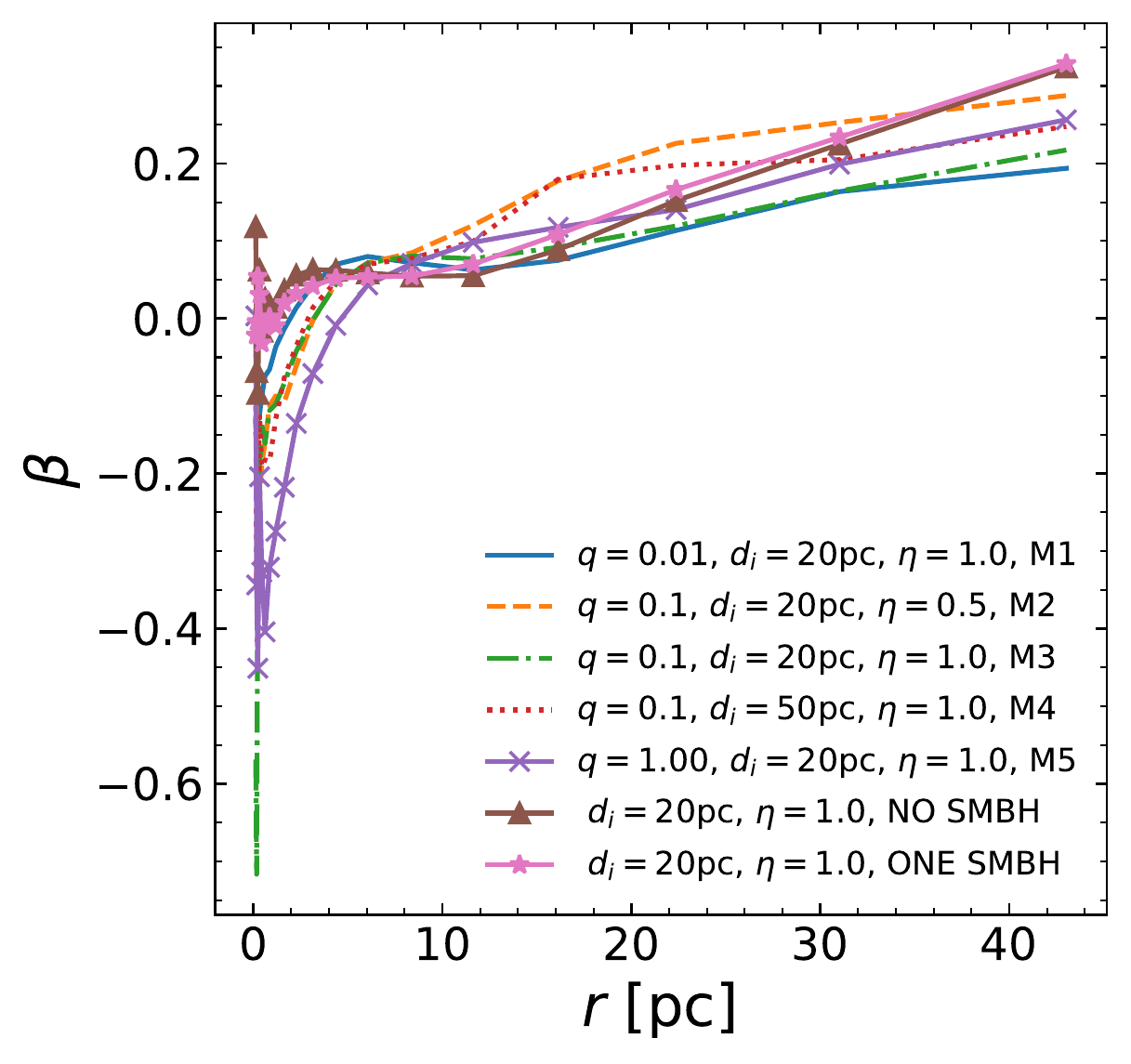}
    \caption{The left panel shows the $|V_{LOS}/\sigma|$ for our five two-SMBHs models and the two additional models with no or one SMBH only. Models in which the rotation is more important have larger values of $|V_{LOS}/\sigma|$. The right panel shows the velocity anisotropy parameter $\beta$ for all the models. All the final NSCs with two SMBHs are tangentially anisotropic in their central regions and become radially anisotropic going at larger distances from the centre. The models with no or one SMBH are centrally isotropic and become radially anisotropic at radii larger than 15\,pc.}
    \label{fig:beta_vsigma}
\end{figure*}

\begin{figure}
    \centering
    \includegraphics[width=0.42\textwidth]{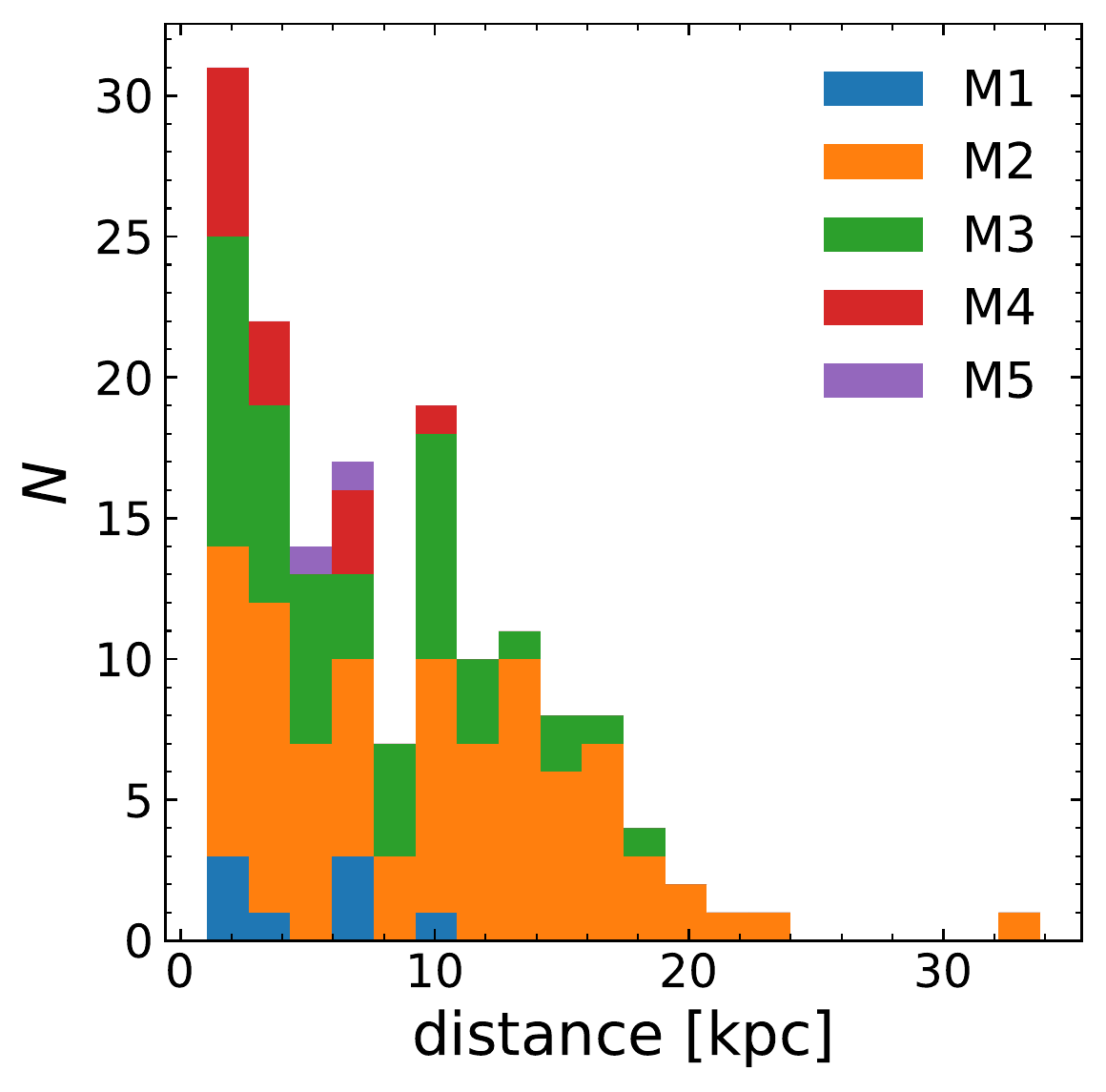}
    \includegraphics[width=0.42\textwidth]{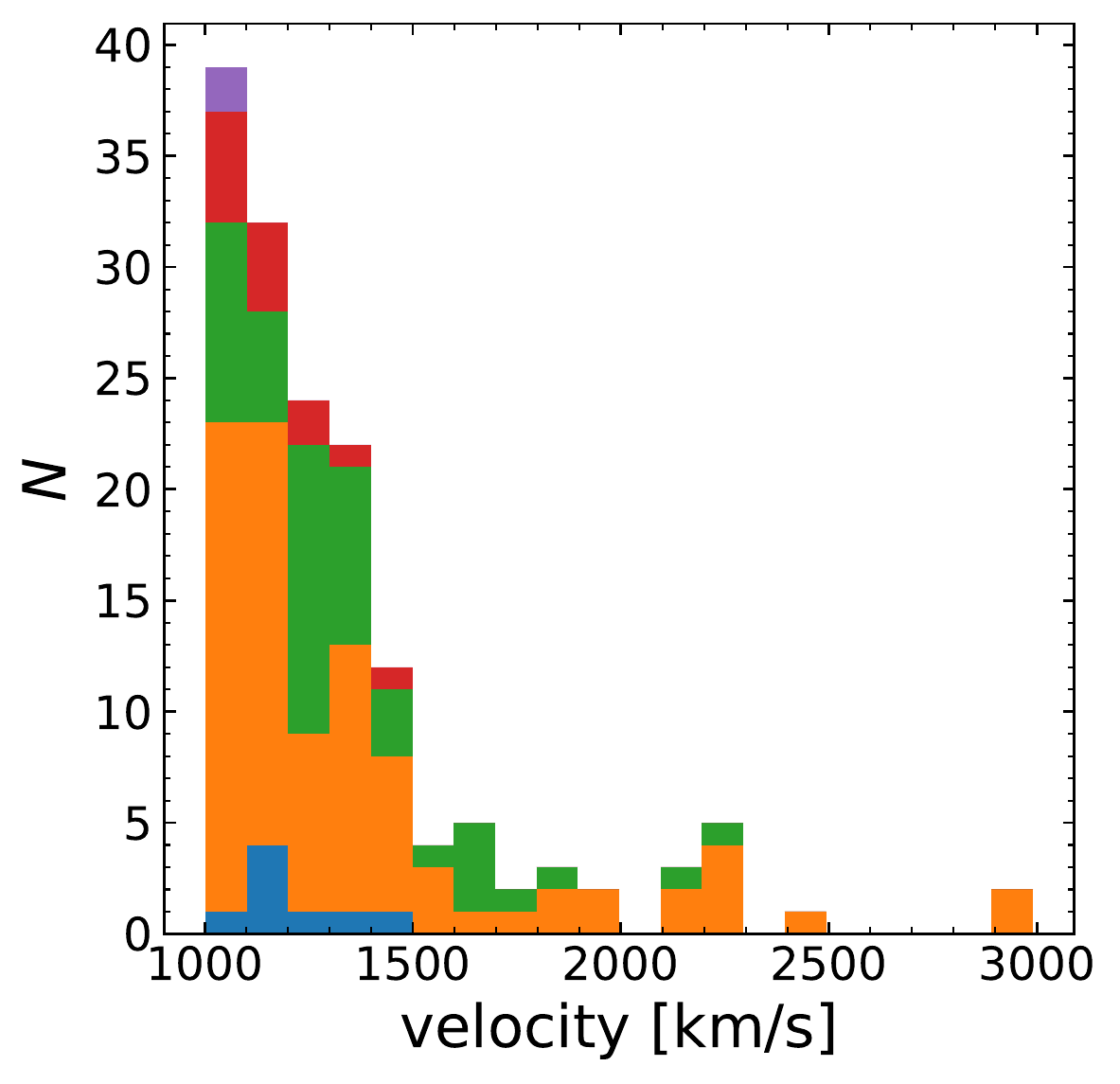}
    \includegraphics[width=0.42\textwidth]{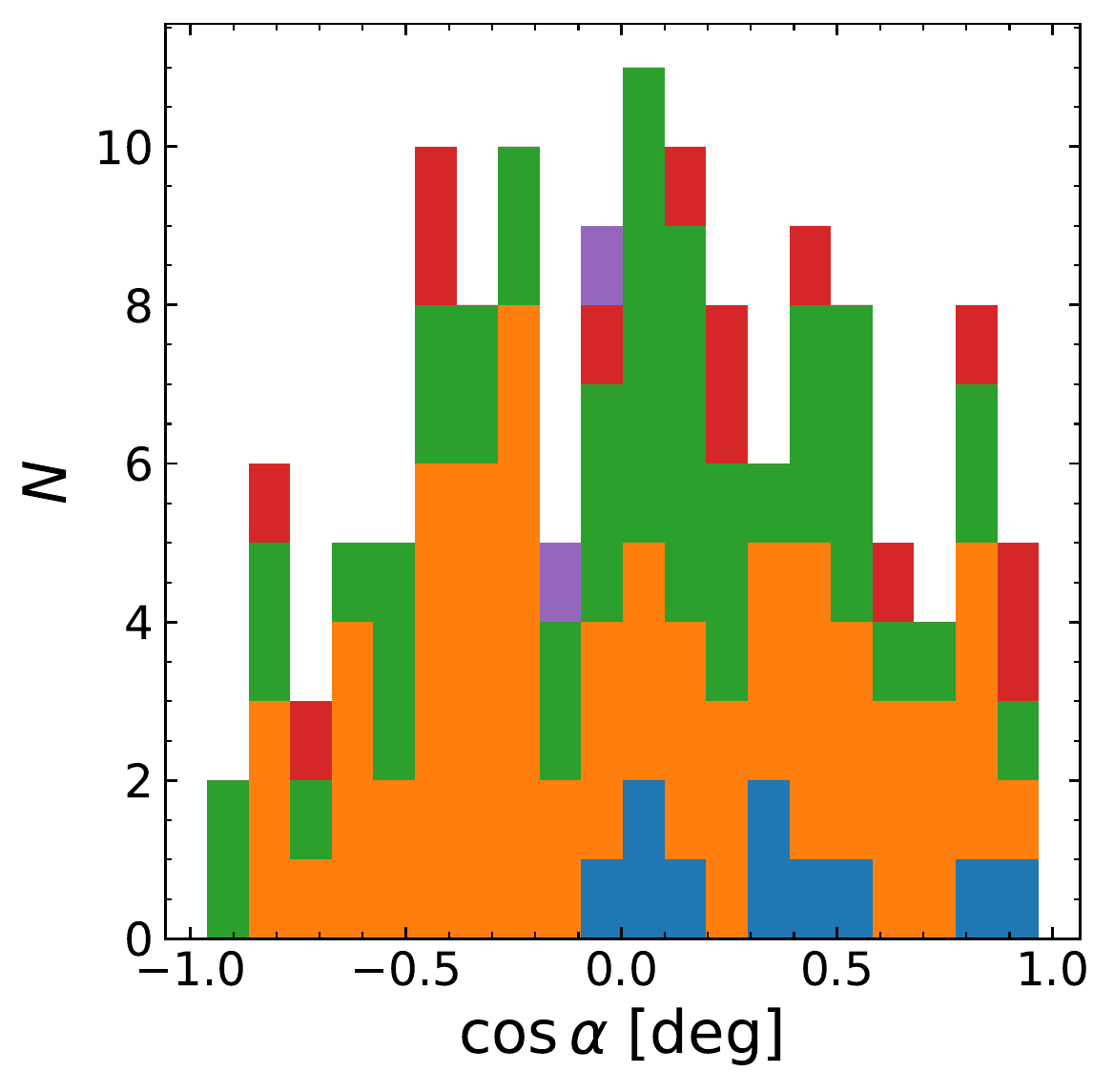}

    \caption{Distribution of the final NSC-centric distances (top panel), velocities (middle panel) and ejection angles (bottom panel) of the hypervelocity stars found in all our two-SMBH models. Different colours are used to differentiate the five models.}
    \label{fig:distr_hvs}
\end{figure}

\begin{figure}
    \centering
    \includegraphics[width=0.42\textwidth]{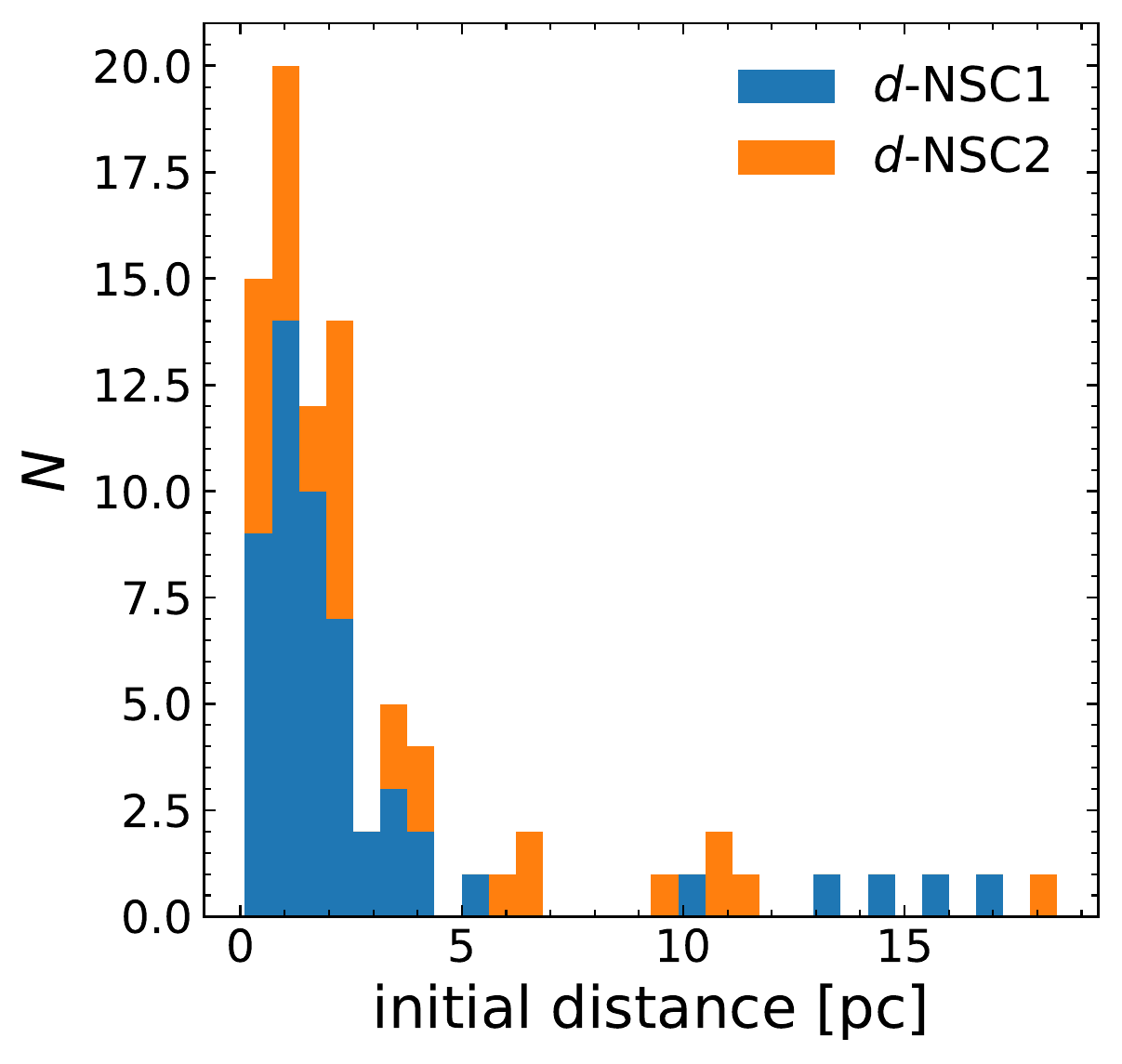}
    \caption{Distribution of the initial NSC-centric distances of the HVSs found in the M2 model.}
    \label{fig:distr_dist_in_M2}
\end{figure}

%\begin{figure}
%    \centering
%    \includegraphics[width=0.50\textwidth]{Figures/hyper_plot_M2.eps}
%    \caption{Spatial distribution of the hypervelocity stars found in the model M1. The direction of the arrow is the direction of the velocity vector of each star and its length is proportional to the value of the velocity. The blue arrow are for stars initiall in NSC1 and the orange arrows are for stars initially in NSC2. The black arrow shows the direction of the total angular momentum of the NSC.}
%    \label{fig:hvs}
%\end{figure}

\subsection{The kinematics of the merger result}
Because of the conservation of the orbital angular momentum of the progenitors, the merger process leaves signatures on the kinematic structure of the final NSC. 
All the results presented in this section refer to the clusters seen edge-on. This choice maximizes the observed rotational signature. However, galaxies are observed at random inclinations, and this needs to be considered when comparing to observations.
As shown by their velocity curves, all the NSCs in our sample show a significant amount of rotation (see Figure \ref{fig:vel_curve}). 
The NSCs formed in mergers with $\eta=1.0$, i.e. with the largest initial relative velocities between the two progenitors, are rotating with a peak velocity approximately equal 20\,km/s, independently of the initial NSC distance and SMBH ratio. The M2 NSC, which forms in the $\eta=0.5$ merger (i.e. with the lowest initial relative velocity between the progenitors), is the slowest rotator in our sample, and it shows a peak velocity of $10\,$km/s.\\
The two stellar populations in each progenitor NSC show slightly different rotational velocities. The population coming from the secondary progenitor NSC is always rotating faster than the one coming from the primary progenitor NSC. As in the case of the flattening, this is due to the fact that the reduced dynamical friction effect on NSC2 makes it retain a larger amount of its initial orbital angular momentum with respect to NSC1. This leads to a faster rotation for the stars initially in NSC2. 
In the M5 case, the two stellar populations show the same rotational velocity. In all our models, the peak velocity is reached at a distance of about $20\,$pc from the cluster centre. This distance is approximately equal to two times the half-mass radius of the NSC. \\
 In the NO SMBH model, NSC1 and NSC2 show the same rotation velocity, with a peak rotation slightly higher than what found for M5 (which also produces two populations rotating at the same speed). With only one SMBH, the stars coming from NSC2 rotate faster than those coming from NSC1, in analogy with what seen for NSCs hosting two SMBHs of different masses.   In all simulations we ran, around 60 per cent of the stars are on retrograde orbits. Retrograde and prograde stars come almost in equal number from the two progenitor NSCs. 
The strength of the rotation signature in each NSC can be better assessed using the $|V_{LOS}/\sigma|$ parameter, that quantifies how important the rotational support is, compared to the disordered motion. M1, M3 and M4 show similar $|V_{LOS}/\sigma|$ behaviours, with a central dip at around 0.6 and values of $|V_{LOS}/\sigma|$ ranging between 0.6 and 1.0 at radii larger than 10pc.
M2, the  case with the two progenitor NSCs with an initial low relative velocity, shows the lowest value of $|V_{LOS}/\sigma|$, which, as opposed to the other cases, is almost constant and approximately equal to $0.4$ at any radius. 
In all the other two-SMBH cases, the central value of $|V_{LOS}/\sigma|$ is $\sim 0.6$. $|V_{LOS}/\sigma|$ increases up to a distance equal to $20\,$pc, reaching values close or larger than unity, depending on the merger conditions. The case with $d_i=50\,$pc has the largest $|V_{LOS}/\sigma|$ values at any radius. While the other cases see a decrease or flattening of the $|V_{LOS}/\sigma|$ value outside 20\,pc, in the M4 model $|V_{LOS}/\sigma|$ continues to increase up to a value equal to $1.4$ which is reached at a galactocentric radius of about $40\,$pc. At larger distances, $|V_{LOS}/\sigma|$ starts to decrease. This is linked to the larger initial distance between the progenitor NSCs. Therefore, $|V_{LOS}/\sigma|$ primarily depends on the initial relative orbits between the progenitor NSCs.\\
The models with no or only one SMBH are more rotationally supported than the models with two SMBHs, as shown by the right panel of Figure \ref{fig:beta_vsigma}. The $|V_{LOS}/\sigma|$ ratios are extremely similar between these two models and higher than what is observed for the two-SMBH models, except for M4. This is again linked to the fact that NSCs with no or only one SMBH are able to retain a larger fraction of their orbital angular momentum. We note an asymmetry between the left and right side of the $|V_{LOS}/\sigma|$ sides, probably linked to the orbital setting of the merger. 
\\
Besides being flattened and rotating, the newly formed NSCs are also anisotropic in  velocity space. The right panel of Figure \ref{fig:beta_vsigma} shows the final radial profile of the anisotropy parameter for our final models. The anisotropy parameter is defined as 
\begin{equation}
    \beta(r)=1-\frac{\sigma_\theta(r)^2+\sigma_\phi(r)^2}{2\sigma_r(r)^2}
\end{equation}
where $\sigma_r$, $\sigma_\theta$ and $\sigma_\phi$ are the components of the velocity dispersion in spherical
coordinates. If the system is isotropic, $\beta$ is equal to zero. If radial orbits are dominant then $\beta>0$, while if the majority of the stars are on tangential orbits $\beta$ takes a negative value. In the limit of all circular orbits, $\beta=-\infty$.
The structure of the initially spherical and isotropic progenitors is modified by the merger and $\beta$ bears witness to the violent past of the new NSC. 
All the two-SMBH systems are tangentially anisotropic in their inner regions, while the NSCs show a mild radial anisotropy within $20\,$pc. This behaviour is due to the fact that the SMBHB scatters away preferentially stars on radial orbits, with an efficiency that decreases with the distance from the SMBHs. The significance of the tangentially biased velocity structure in the NSC centre increases with the mass ratio between the SMBHs, $q$. The radial anisotropy increases at larger radii due to escaping, but still loosely bound, stars. M2 and M4, which produce the most radially anisotropic NSCs, show a steady increase of the radial anisotropy. In M4, $\beta$ increases in the region between 10\,pc and 20\,pc and it flattens at larger radii. M2, instead produces an NSC with an increasing radial anisotropy. The rate of increase is smaller at radii larger than 20\,pc. The merger between NSCs with $q=1.0$, i.e. the model M5, produces the centrally most tangentially anisotropic system. While the value of the central tangential anisotropy mainly depends on $q$, the quantity of external radial anisotropy mostly depends on the initial angular momentum, through $\eta$ and $d_i$.\\
 The NO SMBH and ONE SMBH cases have very similar $\beta$ radial profiles. These models are approximately isotropic within the central $15$ pc (see right panel of Figure \ref{fig:beta_vsigma}). Outside this radius, they both become increasingly radially anisotropic, reaching values higher than what seen for the two-SMBH models.
\\
%\go{Need Figure \ref{fig:vel_maps}? The same information is provided in Figure \ref{fig:vel_curve}.} %yes because this is what observations give
Figure \ref{fig:vel_maps} in Appendix \ref{app1} shows the velocity maps of the two components of the M1 (top row) and M5 (bottom row) NSCs, as well as of the entire NSCs. 
The maps are obtained by applying the Voronoi binning procedure described by \cite{Cappellari03} with a fixed signal-to-noise ratio (S/N) of $15$ in each
bin and can be directly compared to the analogues obtained through integral field units (IFU) observations, e.g. done with MUSE. M1 and M5 are extreme cases in terms of the SMBH ratio. In the M1 case, the stars that were initially in NSC1 rotate significantly slower than the stars that populated NSC2. In the M5 case, the stars coming from the two progenitors rotate at a comparable speed, as also seen in Figure \ref{fig:vel_curve}.\\
 NSC1 and NSC2 rotate at similar or approximately equal rates in the NO SMBH and ONE SMBH runs, as shown in the velocity maps in \ref{fig:COMP_voronoi} of Appendix \ref{app1} (see also Figure \ref{fig:vel_curve}). The degree of rotation is similar to what is found for the fastest rotators that contain an SMBHB.

\subsection{Ejection of hypervelocity stars}
The central SMBHB acts as a source of energy; stars that interact with it can be ejected at high speed from the NSC, becoming hypervelocity stars (HVSs).
We define a hypervelocity star as an $N$-body particle that, at the end of the simulation, has a distance from the NSC centre larger than 1\,kpc and a velocity larger than $1\,000\,$km/s. These conditions imply that the star would be able to escape from the host galaxy, as its velocity is significantly larger than the  escape velocity from the host\footnote{Since our stellar particles have a mass of about $152\,\Mo$, each HVS might trace one or more ejected particles in a real system.}. In the M1 case, we observe a total of 10 HVSs, 8 coming from NSC1 and 2 from NSC2. M2 produces the largest number of HVSs (86) with 53 coming from NSC1 and 33 from NSC2.
The interaction with the SMBHB generates 47 HVSs in M3 (34 from NSC1 and 13 from NSC2). We observe 13 HVSs in the M4 case, 8 coming from NSC1 and 5 from NSC2. Finally, M5 only produces 2 HVSs, both coming from NSC2. 
We note that, while for the M1-M4 cases there is no strong correlation with the final SMBHB separation, the small number of HVSs ejected in the M5 case seems to be linked to the larger orbital separation of the two SMBHs, which is one order of magnitude larger than in all the other cases.
%\go{I guess that the number of HVSs would correlate with the separation between SMBHs. May be interesting to have a look at it.} \\
At $t=20\,$Myr the HVSs produced by the NSCs have distances that range between $1$\,kpc and $30$\,kpc from the centre of their host galaxy (see top panel of Figure \ref{fig:distr_hvs}). M2 and M3 show the widest spatial distributions, while the remaining cases, also due to the lower statistics, show HVS distances smaller than 10\,kpc. 
The velocities of the HVSs are distributed between 1000\,km/s and 3000\,km/s (see middle panel of Figure \ref{fig:distr_hvs}). Only a handful of stars, belonging to the NSC produced in M2 and M3, has velocities larger than 2000\,km/s. The velocity distribution peaks at around $1000$\,km/s and the M1, M4 and M5 cases produce HVSs with velocities smaller than 1500\,km/s.
The cosines of the ejection angles calculated with respect to the SMBHB rotation axis (see bottom panel of \ref{fig:distr_hvs}) are non-uniformly distributed in the range $[-1,1]$. The distribution shows multiple peaks dominated by stars ejected at an angle of $90\,$deg, i.e. along the plane of rotation. %Figure \ref{fig:hvs} shows the velocity vectors of the ejected stars in the M2 case. The direction of the SMBHB angular momentum is also shown as a black arrow. The HVSs seem to be ejected in any direction and can be found at large distances from the galactic centre.
 We inspected the origin of the HVSs in the M2 model, which ejects the largest number of stars (see Figure \ref{fig:distr_dist_in_M2}).  The majority of the ejected stars come from the very central regions of their progenitor NSC; 68 per cent of the HVSs coming from NSC1 were initially inside the central 2\,pc of its progenitor, while 50 per cent of the  HVSs coming from NSC2  were initially within 2\,pc from the centre of their progenitor. Stars initially inside the central few parsecs of their progenitor are more probable to end up at the centre of the final NSC \citep{PMB14}, where they can interact with the binary and be ejected as HVSs.  Moreover, we find that all the HVSs produced in this simulation have been ejected after the SMBHB formation\footnote{The first HVS is ejected after 2.5\,Myr, and the binary forms in less than 2\,Myr.}, while they were orbiting the central parsec of the final NSC. The ejection of stars continues up to the end of the simulation\footnote{The last HVS is ejected at around $19$\,Myr from the beginning of the simulation.}. The large majority ($\sim90$ per cent) of the HVSs found in M2 were within a distance smaller than 0.5pc from the SMBHB before their ejection, and about 30 per cent of them come from a distance smaller than 0.05pc from the central SMBHB.
 No HVSs are ejected in the NO SMBH and ONE SMBH cases, clearly indicating that all the HVSs found in the mergers with two SMBHs are ejected by an encounter with the SMBHB.
 
\subsection{Is there a nuclear stellar disk associated with the NSC?}\label{sec:nsd}
Stars coming from the progenitor NSCs redistribute far beyond their initial distances (see Figure \ref{fig:density}). These stars are scattered at large distances, up to $1\,$kpc, during the merger event and are still bound to the final system. They redistribute in a rotating disc and, in the presence of an external galactic potential, could become part of the nuclear stellar disc (NSD) of the galaxy. 
NSDs are disk structures of few hundred pc size, observed in a wide number of galaxies \cite[see, e.g.][]{Balcells07, Gadotti19, Gadotti20}. Their formation is thought to be linked to gas inflow followed by in situ star formation. The origin of the gas is not yet clear and different hypothesis, including galaxy mergers, have been proposed  and simulated \cite[e.g.][]{Mayer08, Medling14}. Other proposed funnelling processes are nested bars \citep{Shlosman89}, magneto-rotational instability \citep{Milosavljevic04} and  cloud–cloud mergers \citep{Bekki07}. 
Using kinematic data, \cite{Schultheis21} found that the NSD of the Milky Way is kinematically and chemically distinct from the Galactic bulge and from the central NSC. The NSD is more metal-rich than the bulge, and more metal-poor than the central NSC. While stars in the Galactic bulge  are kinematically hot, the NSD shows a kinematically cool and metal rich component, where the velocity dispersion decreases with increasing metallicity, opposite to what is found for the Galactic bulge. These findings are in agreement with \cite{Nogueras20}, who found clearly distinct star formation histories for the bulge, NSD and NSC. Using molecular gas tracers of the central molecular zone, \cite{Schultheis21} found that the gas rotation in the central molecular zone is comparable to the rotation of the NSD metal-rich population. As shown by hydrodynamical simulations \citep[see e.g.][]{Fux99,Li15,Ridley17,sormani2018a, sormani2018b, sormani2019, tress2020}, gas infall at the Galactic centre following the formation of a galactic bar can form a kinematically cold, rotating NSD. Given their dynamical properties, metal-rich stars might have therefore formed from this process. On the other hand, the metal-poor stars of the NSD rotate at a slower rate and show signs of counter-rotation, suggesting that they could have had a different origin. \\
In our simulations, we detect a slow rotation of the stars bound to the NSCs up to radii larger than 100\,pc (see Figure \ref{fig:app_rot} for an example of velocity maps build for the central 200pc$\times$200pc of the M4 model). 
We predict that, in the case of galaxy mergers, part of the NSD might come from the NSCs of the progenitor galaxies. This stellar population would show different chemical, structural and kinematic properties with respect to the in situ component of the NSD. 
The NSC-born stellar population of the NSD must show links to the central NSC (e.g. similar stellar populations, star formation history, rotation direction) and continuity in shape and kinematics. A more detailed and complete study of mergers that takes into account the galactic potential is necessary to improve this results and provide more detailed predictions to be compared with observations \citep[see e.g.][]{Gadotti20}. The nuclear disc components formed through this process would be complementary to those found by \cite{Gadotti20}, which show a bar-driven origin linked to larger scale processes. 

\section{Discussion and conclusions} \label{sec:disc}
%\go{The purpose of the paper is finding observables useful for SMBHB searching. We should comments what observers should look for and discuss the feasibility.} 
Galaxy mergers are common in the Universe and have contributed to the mass growth of galaxies \citep{deBlok10, Newman12, Hill17}. As most of the massive galaxies host a central SMBH, galaxy mergers are a mechanism that can bring to the formation of SMBH binaries. The formation of these systems is more efficient when  the galaxies that merge are nucleated \citep{VanW2014, Biava19, Ogiya20}, i.e. if their SMBH is surrounded by an NSC, a dense and massive stellar system with half-light radii of the order of a few parsecs \citep[see, e.g.][]{Neumayer20}.\\
\cite{Ogiya20} simulated five cases of merger between two NSCs, each hosting a central SMBH. While the mass of the NSC is always equal to $10^7\,\Mo$, the ratio between the SMBH masses is different in each simulation and varies between 0.01 and 1.0. The models adopt different initial distances between the NSCs (either 20\,pc or 50\,pc) and different amounts of initial orbital angular momentum. As a comparison, we have also modelled the merger between two SMBH-less NSCs and the merger between two NSCs, one of which hosts an SMBH and the other does not.\\
%\go{Not sure if we need to summarize our previous paper here.} -> I think the main results are already summarized below
In all the explored cases, the NSCs merge in a few Myr time. The models with no or only one SMBH require more time to merge. On the same timescale, in the two-SMBH cases, the separation between the SMBHs decreases by a few orders of magnitude. This process is particularly efficient because of the combined effect of dynamical friction, stellar hardening, and -- when the two SMBHs are present -- of the deceleration added by the `ouroboros effect', a drag force caused by the stars in the tidal streams of the NSCs.  
When the binary enters the hardening phase, the separation starts to decrease because of the gravitational slingshots between the SMBH binary and stars. The duration of this phase also depends on the SMBH mass ratio. 
All the simulations have been run for a time $t=20\,$Myr. After this time, the binary becomes increasingly hard, slowing down the simulation significantly with a consequent increase in the computational cost.  
The SMBH coalescence time has been then estimated analytically and it ranges between  57.6Myr to 5.3Gyr, depending on the adopted orbital initial conditions and on the SMBH mass ratio. Galaxy mergers have therefore strong implications in the emission of GWs in the band detectable by LISA.\\ 
In the analysed simulations, same age NSCs show different properties depending on the initial conditions adopted for the progenitors and for their relative orbit. While the half-mass radius of the final NSC increases with the mass ratio between the SMBHs, the total mass decreases in function of this same quantity. In all our models, each progenitor NSC contributes to approximately half of the mass of the final NSC contained within $d_i$. All the final NSCs are oblate and significantly flattened throughout their whole radial extent, when observed edge-on. This flattening can be easily detected building luminosity maps of real NSCs. All the systems have similar amount of flattening, with the model with $q=0.1$, $\eta=1.0$ and $d_i=50$\,pc slightly more flattened than the other systems. The two populations coming from the  progenitors show different flattening. The population initially belonging to the NSC hosting the most massive SMBH is always less flattened than the  population initially belonging to the NSC initially hosting the secondary SMBH. This difference is larger within the half-mass radius of the final NSC. This is linked to the smaller effectivity of the dynamical friction on the secondary NSC and on its SMBH, that consequently retains a larger fraction of its initial orbital angular momentum.  The difference in flattening between the two populations is smaller for the system with lower orbital angular momentum, and approximately zero for the system formed from the merger of NSCs hosting same mass SMBHs. 
The central density of the final NSC depends on the merger parameters; smaller mass ratios between the SMBHs give rise to more centrally concentrated NSCs. The central density is less sensitive to the initial distance between the SMBHs.
The cumulative mass radial distribution has a different behaviour depending on the mass ratio and orbital conditions of the merger. Larger mass ratios correspond to stars spread on a larger volume. Models run with no or only one SMBH show the largest cumulative mass at any radius.
The density and cumulative mass behaviour is explained by the increasingly dominant effect of the NSC dynamical contraction over the scattering due to the presence of the SMBHB when the SMBH mass ratio decreases.  Simulations run by \cite{Merritt2001} and \cite{MerrittMilos01} showed that the final density of a system resulting from the accretion of a high-density dwarf galaxy by a low-density giant galaxy strongly depends on the presence or absence of a central SMBH at the centre of the merging galaxies. While the cusp of the dwarf galaxy is disrupted when both galaxies host a central SMBH, if the black hole is removed from the giant galaxy the remnant  acquires a high central density. This happens because, without the action of the SMBH in the giant, the dwarf galaxy is able to more easily retain its initial properties. In our case, removing the secondary SMBH produces a considerably steeper cusp in the final NSC. When no SMBH is present, the final NSC shows a flat core because, in analogy with what found by \cite{Merritt2001} in the absence of SMBHs, the progenitor NSC density profiles are less affected by the merger.
\\
The memory of the merger is imprinted in the kinematic structure of the final NSC. All our final NSCs rotate with velocities between $10\,$km/s and $20\,$km/s. Initial larger relative velocities lead to a stronger rotational signature. We do not observe a strong dependence of the rotational velocity on the mass ratio between the SMBHs. The progenitor populations rotate slightly differently when the mass ratio is different from unity,  with the secondary population rotating faster than the primary again due to its larger ability to retain the initial orbital angular momentum. The strength of the rotation is well traced by the $|V_{LOS}/\sigma|$ parameter. This quantity is the smallest in the case with $\eta=0.5$ and the largest in the case with $q=0.1$, $\eta=1.0$ and $d_i=50$\,pc. There is no clear dependence between  $|V_{LOS}/\sigma|$ and the mass ratio between the SMBHs.  The clusters that start at 50\,pc from each other lose part of their initial mass before arriving at 20\,pc from each other, i.e. the initial distance assumed in the other models. Moreover, the NSCs starting from a larger $d_i$ have less time to relax after the merger, compared to the ones that form from initially closer NSCs. NSCs  that formed in more recent mergers should therefore show a stronger rotational signal when considering the same merger conditions.
The most important parameters that set the rotational pattern of a merged NSC are the relative velocity of the progenitors, the mass of the progenitor NSCs and the age of the final NSC.\\
The merger between two NSCs hosting an SMBH each leaves behind a new NSC that is tangentially anisotropic at its centre. All systems become radially anisotropic outside their half-mass radius. The anisotropy strength depends on the mass ratio between the SMBHs. In particular, larger values of $q$ lead to more tangentially anisotropic NSCs. This effect is due to the fact that an equal mass binary is more efficient in scattering stars away from the central regions of the NSCs, biasing the system towards a larger amount of radial orbits in the external regions, leaving mostly stars on tangential orbits in the central regions. The amount of radial anisotropy in the external regions seems to be independent of the SMBH mass ratio and to depend on a combination of  the values of $d_i$ and $\eta$, i.e. on the initial amount of orbital angular momentum.\\
 We compared the models run with two SMBHs to a merger between two NSCs with no central SMBHs and a merger in which only one of the two NSCs hosts a $10^6\,\Mo$ SMBH. This is useful to understand if, beside finding clues of a past merger, it is also possible to indirectly identify the effect of the presence of an SMBHB. 
The properties of these two additional models are similar to those obtained for other two-SMBH models. However, when no SMBH is involved in the merger, the final system is centrally more spherical than what observed when two equal-mass SMBHs are present in the progenitor NSCs. The system is more compact and massive than what expected for the analogue merger between two NSCs hosting equal mass SMBHs and with the same orbital initial conditions. The two populations also rotate faster than in the equal mass SMBHs case. All these differences are linked to the dynamical friction effectiveness, to the decay time and to the relaxation efficiency, which are all effects that depend on the presence or absence of the SMBHs.\\ 
In more detail, when only an SMBH is present, the main difference with the two-SMBH cases is in the density profile, that shows an extremely steep cusp, reaching values one order of magnitude higher than in all other cases, both in the overall density profile and in the density profile of the population initially hosting the SMBH. This difference is caused by the dominant effect of contraction over scattering when only one SMBH is present in the system \citep{Bahcall76}. The models with only one or no-SMBH are centrally isotropic. They become radially anisotropic outside the central 15\,pc and are more rotationally supported than the cases with two SMBHs. Therefore, looking at the different structural parameters with particular focus on the density profile, the axial ratios and anisotropy radial profile we might be able to distinguish between mergers implying the formation of an SMBHB and other kinds of NSC mergers. We only simulated two comparison models. Cases with smaller SMBHs masses would be  similar to the no SMBH case. Differences due to the orbital parameters will be better explored in a future work.
\\
 We note that the two merging NSCs might have different ages and metallicities.   \cite{Amaro2013} modelled the merger between two or three multimetallic globular clusters, tracking the metallicity of  individual stars in the course of the merger. They found that the cluster resulting from the merger has structural (flattening and rotation) and chemical properties that can be used to trace back their merger origin. In analogy with this study, chemical tagging of stars belonging to an NSC,  together with observations of the dynamical and morphological state of the cluster, can help to shed light on the nature of its progenitors and on the properties of their original host galaxy.\\
SMBHBs are known as sources of HVSs, stars that are able to escape the galaxy potential due to their extremely high velocities. These stars have been used to investigate the central regions of galaxies. We find that in all our simulations, the SMBHB produces HVSs. The model that produces the largest number of HVSs is the one with $q=0.1$, $\eta=0.5$ and $d_i=20$\,pc. Among the systems initially on a relative circular orbit, the model with $q=0.1$ and $d_i=20$\, is the one that leads to the largest number of HVSs. The system with $q=1.0$ is the one that produces the smallest number of HVSs. The HVSs are distributed up to distances of 20\,kpc and are ejected with a range of velocities peaking at about 1000\,km/s and reaching values larger than 2000\,km/s. The ejection angles peak at around $90\,$deg with respect to the angular momentum vector of the SMBHB, i.e. HVSs are launched in the direction of the SMBHB orbital plane, and are not uniformly distributed. The NSC initially hosting the most massive SMBH produces the largest number of HVSs and the number of HVSs strongly depends on the SMBH mass ratio and merger characteristics. In particular, a large SMBHB separation corresponds to a small number of HVSs.  We find that stars ejected as HVSs were typically inside the central few parsecs of their progenitor NSCs at the beginning of the simulation. Stars closer to the central SMBH will, indeed, more probably be delivered closer to the SMBHB in the final NSC with respect to stars initially farther from the respective central SMBH \citep{PMB14}.  No HVSs are ejected in the models with no or only one SMBH.\\
Stars initially belonging to the progenitors are still bound to the final NSC even at distances as large as 1\,kpc. We observe a mild rotation up to 200\,pc from the centre of the final NSC. Velocity substructures are also common at distances equal or larger than 100\,pc. The stars residing in this large scale rotating disc might become part of the NSD of the host galaxy, complementing the in situ component forming at least in part from gas funnelled by the merger.  \\
In conclusion, if an observed NSC shows two stellar populations, possibly with different chemical properties and ages, both rotating, flattened and with a central tangential anisotropy and external radial anisotropy we might suspect the presence of an SMBHB at its centre. 
%No strong rotation and a mild flattening and a different anisotropy pattern is expected in the absence of central SMBHs. This is because if only one or no SMBH is present, we expect the merger to be slower (dynamical friction is less efficient if the central density of the NSC is lower) and less orbital angular momentum would be retained by the merging clusters. Moreover, there would be no binary scattering away stars on central radial orbits, leaving a different imprint on the anisotropy. Finally, without the SMBHs there would be no `ouroboros' effect, substantially modifying the result of the merger.
\\
In all explored cases, the merger is able to leave strong signatures on the small and large scale structure of the newly formed NSC.  The strength of the signatures depend on the mass ratio between the SMBHs and on the orbital conditions of the merger. 
In addition, the presence of HVSs points to an interaction with an SMBHBs, and their number and kinematic properties are a direct consequence of the SMBHB properties. The dynamical structure of observed NSCs could therefore provide clues on the merger origin of the system and on the presence and properties of the central SMBHB. \\
Current instruments such as JWST and future large scale facilities, like the multi-object spectrographs HARMONI and MOSAIC at the ELT will be able to identify stellar populations in galactic nuclei, providing kinematic data and information on the shape and structure of many external NSCs. With these instruments, it will be possible to observe the dynamical effects of a merger, as these extend up to large NSC-centric distances. Spatial resolution will be crucial to inspect the very central regions of NSCs, to clearly detect the effects of the presence of an SMBHB.\\
At our knowledge, this is the first attempt to understand the large-scale dynamical effects of the merger between two NSCs hosting a central SMBH. More systematic and wider studies, exploring a larger parameter space, longer timescales and considering the external potential of the host galaxies will be necessary to provide key information to pinpoint SMBHBs and to infer their properties using more easily accessible large scale photometric and spectroscopic observations of external NSCs.  

\section*{Acknowledgements}
We thank the referee for their valuable comments that improved the paper. AMB acknowledges funding from the European Union’s Horizon 2020 research and innovation programme under the Marie Sk\l{}odowska-Curie grant agreement No 895174. This work has made use of the computational resources obtained through the DARI grant A0120410154. 
GO was supported by the National Key Research and Development Program of China (No. 2022YFA1602903), the Fundamental Research Fund for Chinese Central Universities (Grant No. NZ2020021, No. 226-2022-00216), and the Waterloo Centre for Astrophysics Fellowship. The two new models presented in this paper have been run on the GPU-equipped supercomputer Jean-Zay at IDRIS, the national computing centre for the CNRS (``Centre national de la recherche scientifique'').

%%%%%%%%%%%%%%%%%%%%%%%%%%%%%%%%%%%%%%%%%%%%%%%%%%
\section*{Data Availability}

Simulation data and analysis codes are available upon request.



\bibliographystyle{mnras}
\bibliography{nsc_mergers} % if your bibtex file is called example.bib

%%%%%%%%%%%%%%%%%%%%%%%%%%%%%%%%%%%%%%%%%%%%%%%%%%

%%%%%%%%%%%%%%%%% APPENDICES %%%%%%%%%%%%%%%%%%%%%

\appendix
\section{Rotation inside the NSC} \label{app1}
\begin{figure*}
    \centering
    \includegraphics[width=0.35\textwidth, trim= 0cm 0 0 0 , clip]{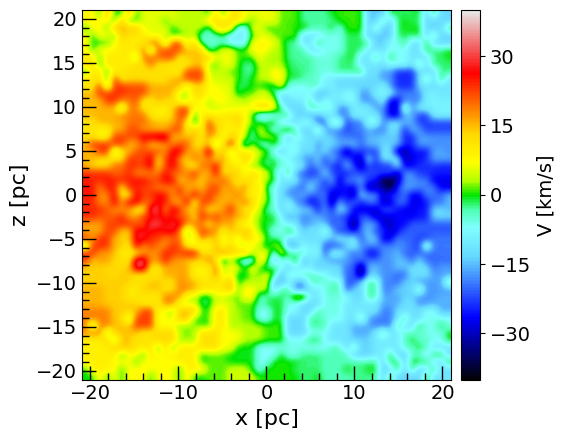}
    \includegraphics[width=0.32\textwidth, trim= 1cm 0 0 0 , clip]{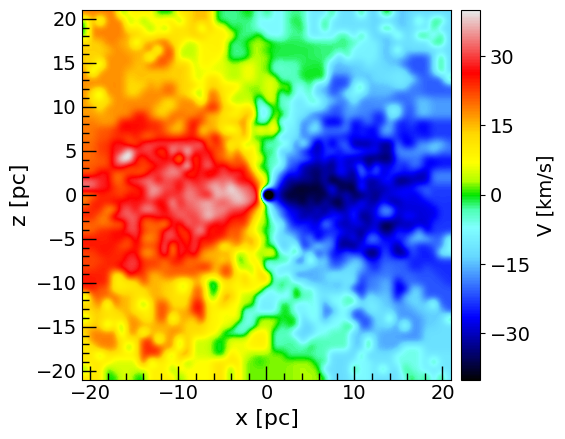}
    \includegraphics[width=0.32\textwidth, trim= 1cm 0 0 0 , clip]{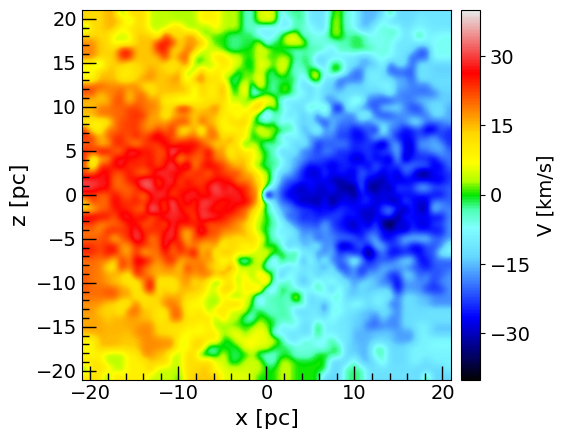}

    \includegraphics[width=0.35\textwidth, trim= 0cm 0 0 0 , clip]{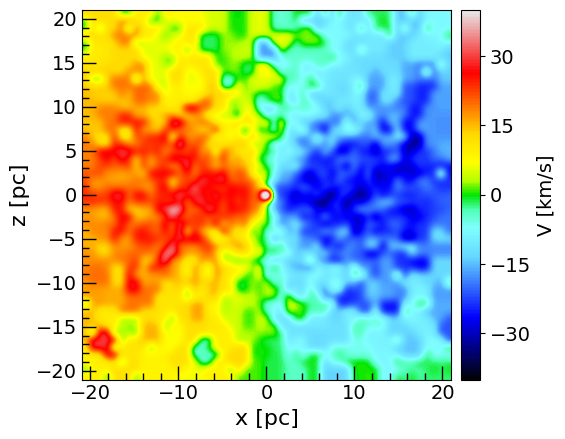}
    \includegraphics[width=0.32\textwidth, trim= 1cm 0 0 0 , clip]{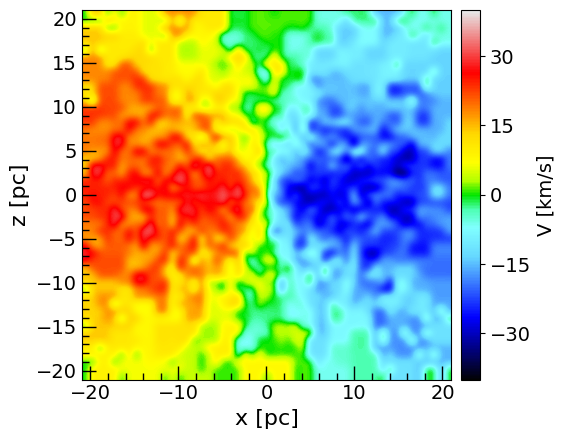}
    \includegraphics[width=0.32\textwidth, trim= 1cm 0 0 0 , clip]{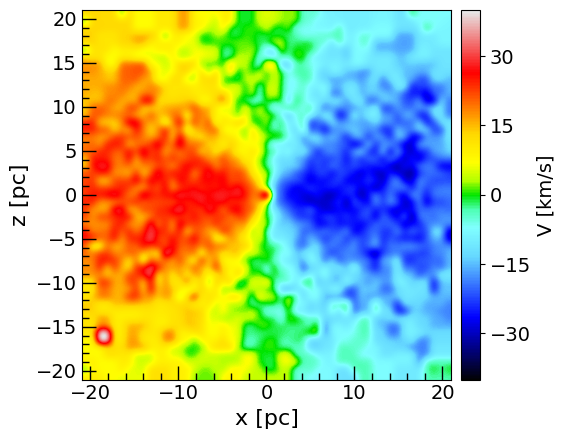}   

    \caption{Velocity maps for the M1 (top panels) and M5 (bottom panels) models. The left panel is for the NSC1 stars, the middle panel is for the NSC2 stars, and the right panel is for the entire system. All the simulated systems rotate at the end of the simulation, and in this plot we are showing the extreme cases in terms of initial conditions of the merger.}
    \label{fig:vel_maps}
\end{figure*}

\begin{figure*}
    \centering
    \includegraphics[width=0.35\textwidth, trim= 0cm 0 0 0 , clip]{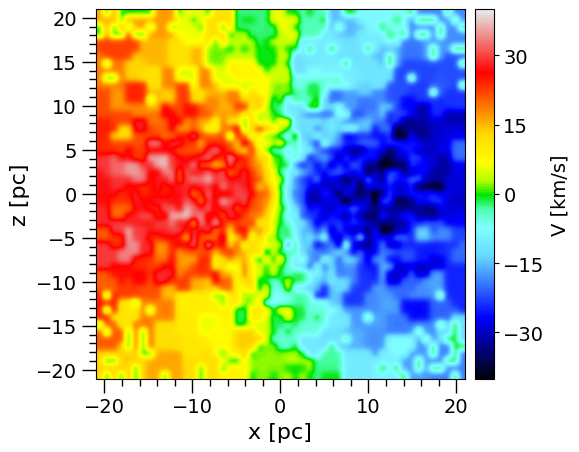}
    \includegraphics[width=0.32\textwidth, trim= 1cm 0 0 0 , clip]{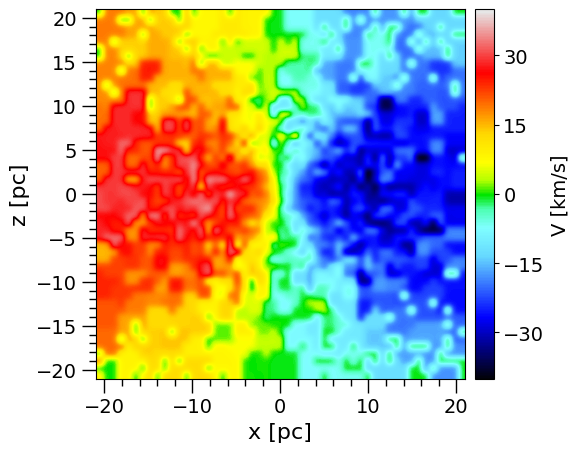}
    \includegraphics[width=0.32\textwidth, trim= 1cm 0 0 0 , clip]{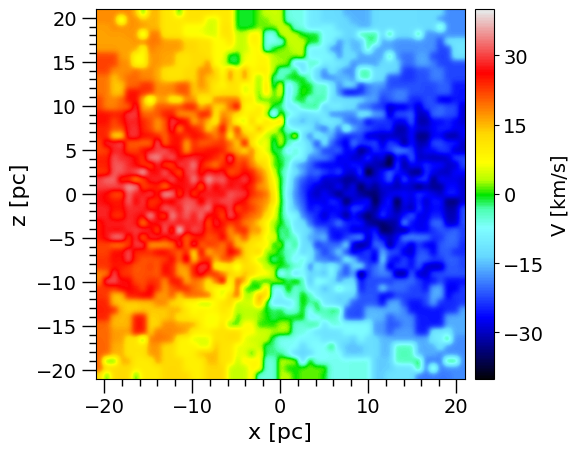}

    \includegraphics[width=0.35\textwidth, trim= 0cm 0 0 0 , clip]{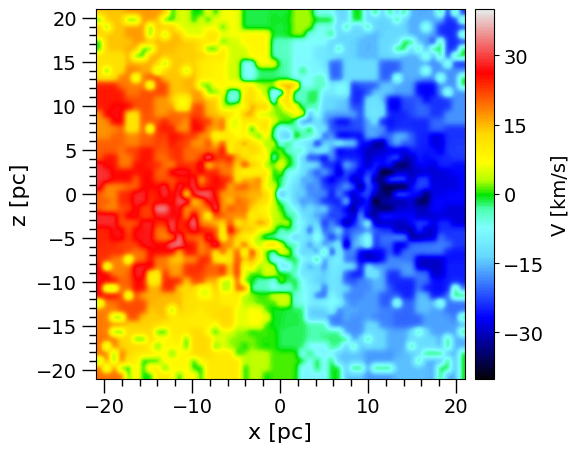}
    \includegraphics[width=0.32\textwidth, trim= 1cm 0 0 0 , clip]{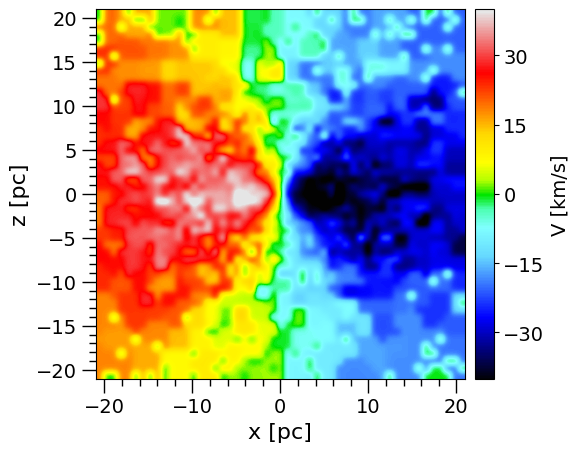}
    \includegraphics[width=0.32\textwidth, trim= 1cm 0 0 0 , clip]{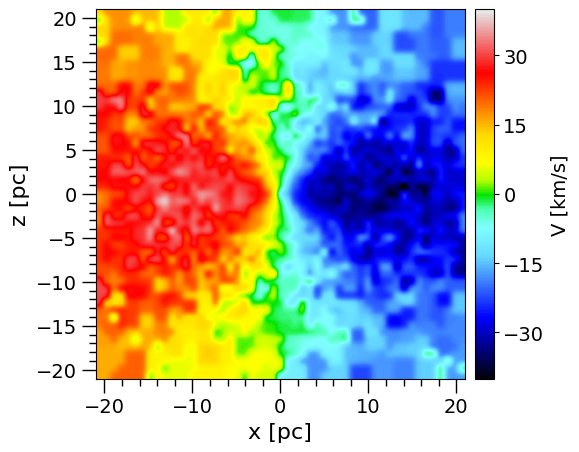}  
    \caption{Velocity maps for the two progenitors of the final NSC (NSC1 on the left panel, NSC2 on the middle panel) and for the entire final NSC (right panel). The top row is for the NO SMBH model and the bottom row is for the ONE SMBH model.}
    \label{fig:COMP_voronoi}
\end{figure*}

\section{Rotation outside the NSC} \label{app2}
\begin{figure}
    \centering
    \includegraphics[width=0.45\textwidth, trim= 0cm 0 0 0 , clip]{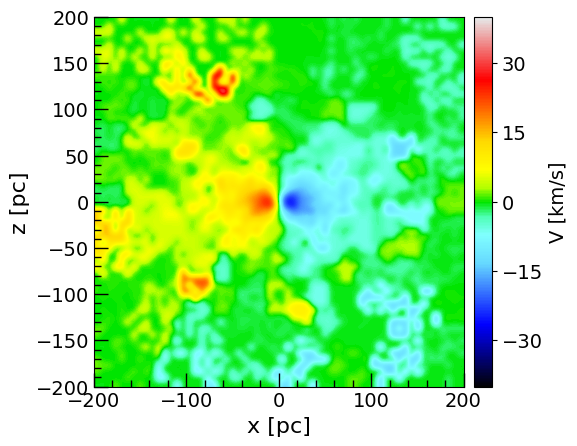}
    \includegraphics[width=0.45\textwidth, trim= 0cm 0 0 0 , clip]{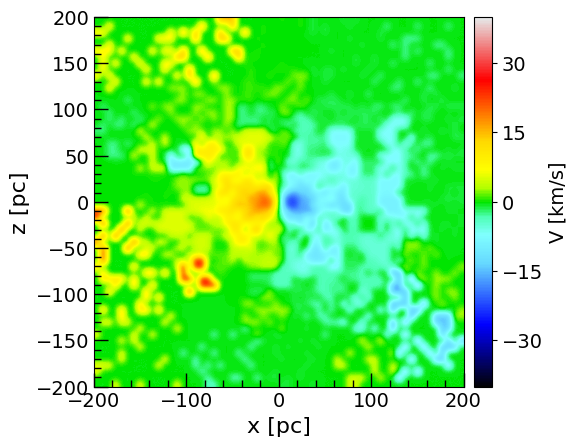}
    \includegraphics[width=0.45\textwidth, trim= 0cm 0 0 0 , clip]{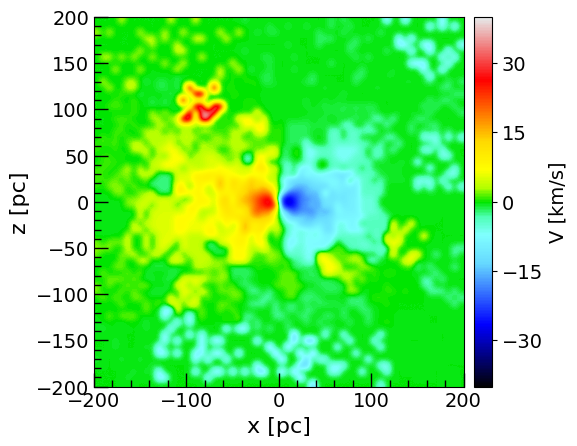}

    \caption{Velocity maps for the model M4, plotted within the central 200$\times$200\,pc$^2$. The top panel is for the entire system, the middle panel for the stars belonging to NSC1, and the bottom panel for the stars belonging to NSC2. The rotation is still visible at radii larger than 100\,pc. }
    \label{fig:app_rot}
\end{figure}

%%%%%%%%%%%%%%%%%%%%%%%%%%%%%%%%%%%%%%%%%%%%%%%%%%

% Don't change these lines
\bsp	% typesetting comment
\label{lastpage}
\end{document}